\documentclass[12pt] {article}
\usepackage{latexsym}{\rm }
\usepackage{graphics}
\usepackage{graphpap}
\newtheorem{lemma}{Lemma}

\newtheorem{proposition}{Proposition}
\newcommand{\beq}{\begin{equation}}
\newcommand{\feq}[1]{\label{#1} \end{equation}}
\newcommand{\beqr}{\begin{eqnarray}}
\newcommand{\feqr}{\end{eqnarray}}
\def\non{\nonumber}
\def\noi{\noindent}

\newcommand{\rf}[1]{(\ref{#1})}

\def\np#1#2#3{Nucl. Phys. {\bf{B#1}} (#2) #3}
\def\cmp#1#2#3{Comm. Math. Phys. {\bf{#1}} (#2) #3}

\def\mpl#1#2#3{Mod. Phys. Lett. {\bf{A#1}} (#2) #3}
\def\jmp#1#2#3{J. Math. Phys. {\bf{#1}} (#2) #3}

\renewcommand{\thefootnote}{\fnsymbol{footnote}}

\begin{document}

\begin{center}


{\Large \bf Boundary Structure and Module Decomposition of the Bosonic $Z_2$ Orbifold Models with $R^2=1/2k$.}\\
[4mm]

\large{Agapitos Hatzinikitas} \\ [5mm]

{\small University of Crete, \\
Department of Applied Mathematics, \\
L. Knosou-Ambelokipi, 71409 Iraklio Crete,\\
Greece, \\
Email: ahatzini@tem.uoc.gr}\\ [5mm]

\large{and} \\ [5mm]

\large{Ioannis Smyrnakis} \\ [5mm]

{\small University of Crete, \\
Department of Applied Mathematics, \\
L. Knosou-Ambelokipi, 71409 Iraklio Crete,\\
Greece, \\
Email: smyrnaki@tem.uoc.gr} \vspace{5mm}

\end{center}

\begin{abstract}

  The $Z_2$ bosonic orbifold models with compactification radius $R^2=1/2k$ are examined in the presence of boundaries.
  Demanding the extended algebra characters to have definite conformal dimension and to consist of an integer sum of
  Virasoro characters, we arrive at the right splitting of the partition function.  This is used to derive
  a free field representation of a complete, consistent set of boundary states, without resorting to a basis
  of the extended algebra Ishibashi states. Finally the modules of the extended symmetry algebra that correspond 
	to the finitely many characters are identified inside the direct sum of Fock modules that constitute the space 
	of states of the theory. 

\end{abstract}
\newpage

\section{Introduction}
\renewcommand{\thefootnote}{\arabic{footnote}}
\setcounter{footnote}{0}
\par The boundary structure of the $Z_2$ bosonic orbifold models with $R^2=1/2k$ is examined.  It was discovered
that for these particular values of the radius the unorbifolded torus models admit a complete consistent set of
Newmann boundary states \cite{ioannis}. To derive the complete set of boundary states for the orbifold
 it is necessary to find the
correct splitting of the partition function of the theories, and from this to read off the characters.  This is achieved by
requiring each character to have definite conformal dimension and to split into a positive integer sum of
Virasoro characters.  The guiding principle is the identification of the $k=1$ orbifold model with the $k=4$ torus model,
which happens at the self-dual point in the Ginsparg classification \cite{ginsp}.

\par The boundary states need to satisfy two consistency conditions.  The first is the Ishibashi condition \cite{ishib},
which is necessary
for the introduction of boundaries in a conformal field theory.  More precisely, it arises from the
requirement that the variation of the correlation functions under a conformal transformation produces the correct
Ward identities.  Of course if the symmetry algebra is larger than the Virasoro algebra, then the 
Virasoro Ishibashi condition
gets modified. The second is the Cardy condition \cite{cardy},
which stems from the identification of the one loop open string amplitude with certain consistent boundary conditions,
to the closed string amplitude between two boundary states that correspond to the open string boundary conditions.
A solution to the Cardy condition for a RCFT was derived in \cite{cardy}, by making use of the Verlinde formula 
\cite{verlinde},
in terms of the S matrix for the characters.  Nevertheless this solution makes use of a complete set of extended
algebra Ishibashi states, a free field representation of which was lacking.

\par Our approach to construct a free field representation of the boundary states consists of
writing the boundary states as linear combinations of the Virasoro Ishibashi states (which are infinite), and then,
instead of trying to find the extended algebra Ishibashi states, apply directly the Cardy condition.  This permits
the determination of the free field representation of the boundary states of our theories.
Finally, the consistency of the boundary states is
checked explicitly, rederiving in this way the fusion rules.

\par Next we study the extended algebra module structure of the space of states of the theory.  First the vacuum 
character is used to find a set of generators of the algebra.  The generating functions of these generators include 
chiral bilocal fields in the sense of \cite{todorov}.  Next a set of primary states is identified. The above generators, 
acting on these 
states produce a set of modules of the extended symmetry algebra.  It is shown that the space of states of the theory 
splits into a direct sum of these modules. Furthermore, traces over these modules produce the correct characters 
(in the right splitting) of the theory.   

\section{Decomposition of the Orbifold Partition Function}
We consider the $Z_2$ orbifold conformal field theory of the free
bosonic field theory compactified on a circle of radius
$R=1/\sqrt{2k}$.  For this choice of radius the unorbifolded
theory has been shown to admit a consistent set of Newmann
boundary states \cite{ioannis}.  The action of our theory is
\beqr
S=\frac{1}{2\pi}\int\partial X\bar{\partial }X
\label{action}
\feqr
where $0\le \sigma \le \beta$ and
is invariant under the $Z_2$ symmetry $g:X\rightarrow -X$.  This corresponds to
$\alpha_n\rightarrow -\alpha_n$, $\bar{\alpha}_n\rightarrow -\bar{\alpha}_n$, $\hat{P}\rightarrow -\hat{P}$ and
$\hat{W}\rightarrow -\hat{W}$, where $\hat{P}$ and $\hat{W}$ are the momentum and winding operators respectively.
Their eigenvalues are $m/R$ and $nR$. This symmetry permits us to
have two kinds of boundary conditions, the untwisted one corresponding to $X(\sigma +\beta,t)=X(\sigma ,t)$ 
and the twisted one corresponding to $X(\sigma +\beta,t)=-X(\sigma ,t)$.
\par The space of states associated to the untwisted
sector is
\beqr
H^U=\oplus_{\stackrel{m,n\in Z}{m> 0}}H^U_{mn}\oplus_{n\ge 0}H^U_{0n}
\label{hilbsum}
\feqr
 where
\beqr
H^U_{mn}=\{\alpha_{-n_1}\cdots \alpha_{-n_j}\bar{\alpha}_{-n_{j+1}}\cdots \bar{\alpha}_{-n_{2l}}(|m,n>+|-m,-n>)\}
 \\ \non
\oplus \{\alpha_{-n_1}\cdots \alpha_{-n_j}\bar{\alpha}_{-n_{j+1}}\cdots \bar{\alpha}_{-n_{2l+1}}(|m,n>-|-m,-n>)\}
\label{uhilb}
\feqr
and $n_i\in Z^+$.
This is the space of states that are invariant under the $Z_2$ symmetry.  The Hilbert space corresponding to the
twisted sector is
\beqr
H^T=\{\alpha_{-n_1}\cdots \alpha_{-n_j}\bar{\alpha}_{-n_{j+1}}\cdots \bar{\alpha}_{-n_{2l}}|1/16,1/16>_0 \} \\ \non
\oplus \{\alpha_{-n_1}\cdots \alpha_{-n_j}\bar{\alpha}_{-n_{j+1}}\cdots \bar{\alpha}_{-n_{2l}}|1/16,1/16>_{\pi R}\}
\label{thilb}
\feqr
where $n_i\in (Z+1/2)^+$.  The states $|1/16,1/16>_0$ and $|1/16,1/16>_{\pi R}$ are primary states of conformal
weight (1/16,1/16) corresponding to the fixed points of the symmetry.  Note that the twisted boundary condition
forces the momentum and winding eigenvalues to be zero, while the only option left for the position operator is to take
values at the orbifold points $0,\pi R$.

\par The untwisted Virasoro generators are
\beqr
L^U_m=\frac{1}{2}\sum_{n\in Z}:\alpha_{m-n}\alpha_n:
\label{uvir}
\feqr
giving rise to the Hamiltonian
\beqr
H^U_{\beta }=\frac{2\pi}{\beta}\left(L_0^U+\bar{L}_0^U - \frac{1}{12}\right),
\label{uham}
\feqr
while the twisted ones are
\beqr
L^T_m=\frac{1}{2}\sum_{n\in Z+1/2}:\alpha_{m-n}\alpha_n:+\frac{1}{16}\delta_{m,0}
\label{tvir}
\feqr
giving rise to the Hamiltonian
\beqr
H^T_{\beta }=\frac{2\pi}{\beta}\left(L_0^T+\bar{L}_0^T-\frac{1}{12}\right).
\label{tham}
\feqr

The radius R orbifold partition function has the well known form \cite{ginsp}
\beqr
Z^R_{orb}(q)=\frac{1}{2}\left(Z^R_{circ}(q)+ \frac{|\theta_2(q)\theta_3(q)|}{\eta^2(q)}
+\frac{|\theta_2(q)\theta_4(q)|}{\eta^2(q)}+\frac{|\theta_3(q)\theta_4(q)|}{\eta^2(q)}\right).
\label{part}
\feqr
Here, $q=e^{2\pi i\tau}=e^{-2\pi \beta }$ where we have taken $\tau $ to be pure imaginary, as in \cite{cardy}.
The precise definitions of the theta and eta functions are given in Appendix A.
In the particular case of $R^2=1/2k$, $k\in Z^+$, the $Z^R_{circ}(q)$ simplifies considerably \cite{ioannis} and takes the
form
\beqr
Z^k_{circ}(q)=\sum_{s=0}^{2k-1}\chi_s(q)\chi_s(q)
\label{circpart}
\feqr
where
\beqr
\chi_s(q)=\sum_{n\in Z}\frac{q^{k(n+\frac{s}{2k})^2}}{\eta(q)}.
\label{cchar}
\feqr
It is convenient to define the following extra chiral components:
\beqr
\chi_a(q)&=&\frac{1}{2}\frac{\sqrt{\theta_2(q)\theta_3(q)}}{\eta(q)}=
\frac{1}{\sqrt{2}}\frac{q^{1/48}}{\prod_{n=1}^{\infty }(1-q^{n-1/2})}=
\frac{1}{2\sqrt{2}}\frac{\sum_{n\in Z}q^{\frac{1}{4}(n-1/2)^2}}{\eta (q)}, \\ \non
\chi_b(q)&=&\frac{1}{2}\frac{\sqrt{\theta_2(q)\theta_4(q)}}{\eta(q)}=
\frac{1}{\sqrt{2}}\frac{q^{1/48}}{\prod_{n=1}^{\infty }(1+q^{n-1/2})}=
\frac{1}{2}\frac{\sum_{n\in Z}q^{\frac{1}{4}(n-1/2)^2} e^{i\frac{\pi}{2}(n-1/2)}}{\eta (q)}, \\ \non
\chi_c(q)&=&\frac{\sqrt{\theta_3(q)\theta_4(q)}}{\eta(q)}=
\frac{q^{-1/24}}{\prod_{n=1}^{\infty }(1+q^{n})}=
\frac{\sum_{n\in Z}q^{n^2}(-1)^n}{\eta (q)}.
\label{chiabc}
\feqr
Using this notation the orbifold partition function becomes
\beqr
Z^k_{orb}=\frac{1}{2}\sum_{s=0}^{2k-1}\chi_s(q)\chi_s(q)+2\chi_a(q)\chi_a(q)+2\chi_b(q)\chi_b(q)+
\frac{1}{2}\chi_c(q)\chi_c(q).
\label{opart1}
\feqr
Noticing that $\chi_s(q)=\chi_{2k-s}(q)$ we can rewrite the above sum as
\beqr
Z^k_{orb}&=&\frac{1}{2}\chi_0(q)\chi_0(q)+\sum_{s=0}^{k-1}\chi_s(q)\chi_s(q)+\frac{1}{2}\chi_k(q)\chi_k(q)\\ \non
&+& 2\chi_a(q)\chi_a(q)+2\chi_b(q)\chi_b(q)+\frac{1}{2}\chi_c(q)\chi_c(q).
\label{opart2}
\feqr

\par One problem in this decomposition of the partition function is that $\chi_a(q)$ and $\chi_b(q)$ do not
have a definite conformal dimension. To eliminate this problem we rewrite
$2\chi_a(q)\chi_a(q)+2\chi_b(q)\chi_b(q)$ in the form $2\chi_+(q)\chi_+(q)+2\chi_-(q)\chi_-(q)$
where
\beqr
\chi_+(q)&=&\frac{1}{\sqrt{2}}(\chi_a(q)+\chi_b(q))=\sum_{n\in Z}\frac{q^{4(n+1/8)^2}}{\eta (q)}, \\ \non
\chi_-(q)&=&\frac{1}{\sqrt{2}}(\chi_a(q)-\chi_b(q))=\sum_{n\in Z}\frac{q^{4(n+3/8)^2}}{\eta (q)}.
\label{tchr1}
\feqr
The components \rf{opart2} satisfy
\beqr
\chi_+(e^{2\pi i}q)&=&e^{2\pi i\frac{1}{16}}\chi_+(q), \\ \non
\chi_-(e^{2\pi i}q)&=&e^{2\pi i\frac{9}{16}}\chi_-(q),
\label{transf1}
\feqr
so their conformal dimensions are $1/16$ and $9/16$ respectively.

\par Another problem with this decomposition is that not all components are an integer sum of Virasoro characters.
This is necessary since we are going to search for a larger algebra that groups the infinitely many Virasoro characters 
into finitely many
groups which are to be interpreted as the characters of the larger algebra.  This is because we are searching for an
algebra that is going to contain the Virasoro algebra as a subalgebra of its universal enveloping algebra.  
This suggests that we should change
again the
splitting of the partition function.  To see which is the right splitting it is instructive to look more closely at the
$k=1$ model.  This is none other than the self-dual point in the Ginsparg's classification \cite{ginsp}.
This model should be
the same as the $k=4$ torus model, so we should be able to identify the components of the partition functions.
For the $k=4$ torus model we have the characters
\beqr
\chi^{k=4}_s(q)=\sum_{n\in Z}\frac{q^{4(n+\frac{s}{8})^2}}{\eta(q)} ,\qquad 0\le s\le 7
\label{k4char}
\feqr
while for the $k=1$ orbifold model we have the components
\beqr
\chi^{k=1}_s(q)=\sum_{n\in Z}\frac{q^{(n+\frac{s}{2})^2}}{\eta(q)} ,\qquad 0\le s\le 1
\label{k1orb}
\feqr
as well as $\chi_+(q)$, $\chi_-(q)$ and $\chi_c(q)$.  The identification proceeds as follows:
\beqr
\frac{1}{2}\left(\chi^{k=1}_0(q)+\chi_c(q)\right)=\chi^{k=4}_0(q), \\ \non
\frac{1}{2}\left(\chi^{k=1}_0(q)-\chi_c(q)\right)=\chi^{k=4}_4(q), \\ \non
\frac{1}{2}\chi^{k=1}_1(q)=\chi^{k=4}_2(q)=\chi^{k=4}_6(q), \\ \non
\chi_+(q)=\chi^{k=4}_1(q)=\chi^{k=4}_7(q), \\ \non
\chi_-(q)=\chi^{k=4}_3(q)=\chi^{k=4}_5(q).
\label{ident}
\feqr
Under this identification the two partition functions are equal.  So in the case of $k=1$ we have the following
splitting of the orbifold partition function into characters:
\beqr
Z^{k=1}_{orb}=\chi^{* k=1}_+(q)\chi^{* k=1}_+(q)+\chi^{* k=1}_-(q)\chi^{* k=1}_-(q)+2\chi^*_{1}(q)\chi^*_{1}(q)+
2\chi_+(q)\chi_+(q)+2\chi_-(q)\chi_-(q)
\label{opart3}
\feqr
where
\beqr
\chi^{* k=1}_\pm (q)=\frac{1}{2}\left(\chi_0(q)\pm \chi_c(q)\right), \\ \non
\chi^*_1(q)=\frac{1}{2}\chi_1(q)=\sum_{n\in Z}\frac{q^{4k(n+1/4)^2}}{\eta(q)}=
\sum_{n=0}^{\infty }\frac{q^{k(n+1/2)^2}}{\eta(q)}.
\label{newchar}
\feqr

\par This seems to suggest that for general k the right splitting is
\beqr
Z^k_{orb}&=&\chi^*_+(q)\chi^*_+(q)+\chi^*_-(q)\chi^*_-(q)+2\chi^*_k(q)\chi^*_k(q)+\sum_{s=1}^{k-1}\chi_s(q)\chi_s(q) \\ \non
&+& 2\chi_+(q)\chi_+(q)+2\chi_-(q)\chi_-(q).
\label{opart4}
\feqr
Now we have the following lemma:
\begin{lemma}
The extended algebra characters $\chi^*_\pm (q)$, $\chi^*_k(q)$, $\chi_s(q)$ and $\chi_\pm (q)$ split into an integer sum
of Virasoro characters.
\end{lemma}
The proof of this lemma is in appendix A.
It is worth noting that the splitting into Virasoro characters is different in the cases $k\ne l^2$ and $k=l^2$.
This suggests different behavior of the theories in the two cases. Indeed the models with $k=l^2$ can be identified
with the $D_l$ dihedral group orbifolds of the $SU(2)\times SU(2)$ level one theory, which corresponds to the $k=1$
($R=1/\sqrt{2}$) torus model.

\section{Boundary States}
There are two conditions that the boundary states have to satisfy.  One is the Ishibashi \cite{ishib} condition
and the other is the Cardy \cite{cardy} condition.

\par The Ishibashi condition appears when we restrict the conformal field theory to the upper half plane.  The
variation of the correlation functions under $z\rightarrow z+a(z)$ is given by
\beqr
\! \! \! \! \delta<\phi_{h_1}(z_1,\bar{z}_1)\cdots \phi_{h_N}(z_N,\bar{z}_N)>
&=& \frac{1}{2\pi i}\oint_Ca(z)<T(z)\phi_{h_1}(z_1,\bar{z}_1)\cdots \phi_{h_N}(z_N,\bar{z}_N)>dz \\ \non
&-& \frac{1}{2\pi i}\oint_Ca(\bar{z})<\bar{T}(\bar{z})\phi_{h_1}(z_1,\bar{z}_1)\cdots \phi_{h_N}(z_N,\bar{z}_N)>d\bar{z}
\label{variat}
\feqr
where the contour C contains all the points $z_i$.  This can be deformed to a contour on a large semicircle
and a contour on the real line. The integral over the real line has to vanish for the Ward identities to be valid,
and this implies that $T(z)=\bar{T}(\bar{z})$ on the real line.  Since the real line is the boundary of the
upper half plane this translates (in the closed string picture) into the condition
\beqr
(L_{n}-\bar{L}_{-n})|B>=0
\label{ishcond}
\feqr
for the boundary states $|B>$.
It is worth mentioning that if we have a larger symmetry algebra of generators
$(W_n^{(r)},\bar{W}_n^{(r)})$ then the corresponding Ishibashi condition becomes
\beqr
(W_n^{(r)}-(-1)^s\bar{W}_{-n}^{(r)})|B>=0
\label{gishib}
\feqr
where s is the spin of the generators of the algebra.
\par The solution of the Virasoro Ishibashi condition in the case of a free field with untwisted boundary
conditions gives two kinds of states.  One corresponding to Dirichlet and the other to Newmann boundary conditions
in the target space:
\beqr
|i^D_{m/2R}>&=&e^{\sum_{n=1}^{\infty }\frac{a_{-n}\bar{a}_{-n}}{n}}|m,0>, \\ \non
|i^N_{nR}>&=&e^{-\sum_{n=1}^{\infty }\frac{a_{-n}\bar{a}_{-n}}{n}}|0,n>.
\label{ishibstat}
\feqr
Upon orbifolding only the linear combinations invariant under $Z_2$ survive so we get
\beqr
|i^{UD}_{m/2R}>&=&e^{\sum_{n=1}^{\infty }\frac{a_{-n}\bar{a}_{-n}}{n}}\frac{1}{\sqrt{2}}(|m,0>+|-m,0>), \\ \non
|i^{UN}_{nR}>&=&e^{-\sum_{n=1}^{\infty }\frac{a_{-n}\bar{a}_{-n}}{n}}\frac{1}{\sqrt{2}}(|0,n>+|0,-n>).
\label{circish}
\feqr
In the case of the twisted boundary conditions we get \cite{affleck}:
\beqr
|i^{TD}_{0,\pi R}>&=&e^{\sum_{n=1}^{\infty }\frac{a_{-n+1/2}\bar{a}_{-n+1/2}}{n-1/2}}|1/16,1/16>_{0,\pi R}, \\ \non
|i^{TN}_{0}>&=&e^{-\sum_{n=1}^{\infty }\frac{a_{-n+1/2}\bar{a}_{-n+1/2}}{n-1/2}}\frac{1}{\sqrt{2}}
(|1/16,1/16>_0+|1/16,1/16>_{\pi R}), \\ \non
|i^{TN}_{\pi /2R}>&=&e^{-\sum_{n=1}^{\infty }\frac{a_{-n+1/2}\bar{a}_{-n+1/2}}{n-1/2}}\frac{1}{\sqrt{2}}
(|1/16,1/16>_0-|1/16,1/16>_{\pi R}).
\label{orbish}
\feqr

\par The Cardy condition comes from the identification of the partition function of the closed string between two
boundary states, with the one loop partition function of the open string with the boundary conditions corresponding
to the boundary states.  Here we are going to view the above cylinder as a strip of length $\beta $ and width $1/2$.
The two ends corresponding to length 0 and $\beta $ are identified.  The length corresponds to the open string time,
while the width corresponds to the closed string time.
In the open string picture the partition function (for a general CFT) takes the form
\beqr
Z_{ab}=\sum_in_{ab}^i\chi_i(q)
\label{gopart}
\feqr
where $n_{ab}^i$ counts how many copies of the channel i exist in the open string partition function with boundary
conditions $a$ and $b$.
In the closed string picture we have
\beqr
Z_{ab}=\sum_j<a|j><j|b>\chi_j(\tilde{q})
\label{gcpart}
\feqr
for the boundary states corresponding to the open string boundary conditions.
Note that the time in the open string picture corresponds to
space in the closed string one, and this explains why in the closed string partition function we have
$\tilde{q}=e^{-2\pi /\beta }$ instead of $q=e^{-2\pi \beta}$.
Here $|j>$ is a complete set of states satisfying the extended algebra Ishibashi
condition.  They are normalized according to
\beqr
<j|j'>=\delta_{jj'}S_{0j}
\label{norm}
\feqr
and satisfy
\beqr
<j'|\tilde{q}^{\frac{1}{2}(L_0+\bar{L}_0-\frac{c}{12})}|j>=\delta_{jj'}\chi_j(\tilde{q}).
\label{extish}
\feqr
These are known to be in one to one correspondence with the characters of the theory \cite{petkova}.
The boundary states are constructed as linear combinations of these states.

\par When we identify \rf{gopart} with \rf{gcpart} we get 
\beqr
\sum_iS_i^jn^i_{ab}=<a|j><j|b>
\label{cardy}
\feqr
where $S_i^j$ is the S matrix for the characters.
This is the Cardy condition.  A solution to this condition, once one knows the set of characters and the complete
basis of Ishibashi states, is obtained by using the Verlinde formula \cite{verlinde}.
This formula tells us that
\beqr
\sum_iS_i^jN^i_{kl}=S^j_kS^j_l/S_0^j
\label{verlin}
\feqr
where $N^i_{kl}$ represent the fusion rules.
This suggests the following solution for the Cardy condition \cite{cardy}:
\beqr
|l>&=&\sum_j\frac{S_l^j}{(S_0^j)^{1/2}}|j>, \\ \non
|l^\vee >&=&\sum_j\frac{(S_l^j)^*}{(S_0^j)^{1/2}}|j>, \\ \non
n_{k^\vee l}^i&=&N_{kl}^i.
\label{cardysol}
\feqr
Note that among the boundary states there is a special state, the vacuum sate
\beqr
|0>=\sum_j(S_0^j)^{1/2}|j>
\label{vcst}
\feqr
which satisfies
\beqr
|0^\vee >=|0>, \qquad n^i_{00}=N^i_{00}=\delta^i_0, \qquad n^i_{0l}=N^i_{0l}=\delta^i_l.
\label{vaccond}
\feqr
This means that
\beqr
Z_{00}=\chi_0(q), \qquad Z_{0l}=\chi_l(q).
\label{vacpart}
\feqr
A set of boundary states is {\it consistent} if it contains
a vacuum boundary state satisfying \rf{vacpart} and for different combinations of boundary states
the partition function is an integer sum of characters.
At this point we should remark that if we knew what the states $|j>$ were, the construction of the
boundary states would be easy, since we could read the S matrix from the characters that
appear in the right diagonalization of the torus partition function.  But to construct these states in the
way presented in \cite{ishib}, we need to understand the representation theory of the extended symmetry algebra.
An alternative way to proceed is to construct directly the
boundary states as (infinite) linear combinations of the free field solutions to the Virasoro Ishibashi condition.  After
all, this free field representation of the boundary states makes calculations a lot more tractable.
\par In the sequel we need the inner products of the Virasoro Ishibashi states for the orbifold.  
In the untwisted sector these are given by
\beqr
\label{uprod}
<i^{UN}_{nR}|e^{-H_{\beta }^{U} /2}|i^{UN}_{mR}> &=& \frac{\tilde{q}^{\frac{R^2n^2}{2}}}{\eta (\tilde{q})}\delta_{n,m}, 
\\ \non 
<i^{UD}_{n/2R}|e^{-H_{\beta}^{U} /2}|i^{UD}_{m/2R}> &=& \frac{\tilde{q}^{\frac{n^2}{8R^2}}}{\eta (\tilde{q})}\delta_{n,m},
 \\ \non 
<i^{UN}_{nR}|e^{-H_{\beta}^{U} /2}|i^{UD}_{m/2R}> &=&
\tilde{q}^{-\frac{1}{24}}\prod_{l=1}^{\infty }\frac{1}{1+\tilde{q}^l}\delta_{n,0}\delta_{m,0}
=\frac{\sum_{l\in Z}(-1)^l\tilde{q}^{l^2}}{\eta (\tilde{q})}\delta_{n,0}\delta_{m,0} \\ \non 
&=& \chi_c(\tilde{q})= \sqrt{2}(\chi_+(q)+\chi_-(q))  
\feqr 
 while in the twisted one they are 
\beqr
\label{tprod}
<i^{TN}_{0}|e^{-H_{\beta }^{T} /2}|i^{TN}_{0}>&=&<i^{TN}_{\pi /2R}|e^{-H_{\beta }^{T} /2}|i^{TN}_{\pi /2R}> \\ \non
=<i^{TD}_{0}|e^{-H_{\beta}^{T} /2}|i^{TD}_{0}>&=& <i^{TD}_{\pi R}|e^{-H_{\beta}^{T} /2}|i^{TD}_{\pi R}> \\ \non
&=& \tilde{q}^{\frac{1}{48}}\prod_{l=1}^{\infty }\frac{1}{1-\tilde{q}^{l-1/2}}=
\frac{1}{2}\frac{\sum_{l\in Z}\tilde{q}^{\frac{1}{4}(l-1/2)^2}}{\eta(\tilde{q})} \\ \non
&=&\sqrt{2}\chi_a(\tilde{q})=\frac{1}{\sqrt{2}}(\chi_+^*(q)-\chi_-^*(q)), \\ \non
<i^{TD}_{0,\pi R}|e^{-H_{\beta}^{T} /2}|i^{TN}_{0}>&=&<i^{TD}_{0}|e^{-H_{\beta }/2}|i^{TN}_{\pi /2R}>=
\frac{1}{2}\frac{\sum_{l\in Z}\tilde{q}^{\frac{1}{4}(l-1/2)^2}e^{\frac{\pi i}{2}(n-1/2)}}{\eta(\tilde{q})} \\ \non
&=&\chi_b(\tilde{q})=\frac{1}{\sqrt{2}}(\chi_+(q)-\chi_-(q)), \\ \non
<i^{TD}_{\pi R}|e^{-H_{\beta}^{T} /2}|i^{TN}_{\pi /2R}>&=&
-\frac{1}{2}\frac{\sum_{l\in Z}\tilde{q}^{\frac{1}{4}(l-1/2)^2}e^{\frac{\pi i}{2}(n-1/2)}}{\eta(\tilde{q})}=
-\frac{1}{\sqrt{2}}(\chi_+(q)-\chi_-(q)).
\feqr
Here we have used the identity
\beqr
\label{poisson}
\sum_{n\in Z}\frac{\tilde{q}^{A(n+b)^2}}{\eta (\tilde{q})}= \frac{1}{\sqrt{2A}}\sum_{n\in Z} 
\frac{q^{\frac{n^2}{4A}}e^{2\pi i n b}}{\eta(q)}
\feqr
which comes from the Poisson resummation formula.
\par The next step is to search for the vacuum boundary state in the case $R^2=1/2k$.
This has to be a linear combination of the
twisted and the untwisted Virasoro Ishibashi states and it has to give $Z_{00}(q)=\chi_+^*(q)$.  The reason
is that the character $\chi_+^*(q)$ is the only one that admits the interpretation of a vacuum character since
it has the right conformal dimension (while the character $\chi_-^*(q)$ has conformal dimension 1).
Using the inner products \rf{uprod}, \rf{tprod} it is not very difficult to construct such a state:
\beqr
|X_0^{N+}>=\frac{1}{\sqrt{2}\sqrt[4]{2k}}\sum_{n\in Z}|i^{UN}_{\frac{n}{\sqrt{2k}}}>+\frac{1}{\sqrt[4]{2}}|i^{TN}_0>.
\label{vacstat}
\feqr
\par Next we need to construct boundary states whose partition function with the vacuum boundary state gives all the characters.
A set of such states is
\beqr
|X_0^{N-}>&=&\frac{1}{\sqrt{2}\sqrt[4]{2k}}\sum_{n\in Z}|i^{UN}_{\frac{n}{\sqrt{2k}}}>-\frac{1}{\sqrt[4]{2}}|i^{TN}_0>,
 \\ \non
|X_m^N>&=&\frac{\sqrt{2}}{\sqrt[4]{2k}}\sum_{l=0}^{2k-1}e^{\frac{2\pi ilm}{2k}}\sum_{n\in Z}
|i^{UN}_{\sqrt{2k}(n+\frac{l}{2k})}>, \qquad 1\le m\le k-1 \\ \non
|X_k^{N\pm }>&=&\frac{1}{\sqrt{2}\sqrt[4]{2k}}\sum_{n\in Z}(-1)^n|i^{UN}_{\frac{n}{\sqrt{2k}}}>\pm
\frac{1}{\sqrt[4]{2}}|i^{TN}_{\frac{\pi }{2R}}>, \\ \non
|X_0^{D\pm }>&=&\frac{1}{\sqrt{2}}\sqrt[4]{\frac{k}{2}}\sum_{n\in Z}|i^{UD}_{n\sqrt{\frac{k}{2}}}>
\pm \frac{1}{\sqrt[4]{2}}|i^{TD}_0>, \\ \non
|X_k^{D\pm }>&=&\frac{1}{\sqrt{2}}\sqrt[4]{\frac{k}{2}}\sum_{n\in Z}(-1)^n|i^{UD}_{n\sqrt{\frac{k}{2}}}>
\pm \frac{1}{\sqrt[4]{2}}|i^{TD}_{\pi R}>.
\label{boundstat}
\feqr
These give that
\beqr
Z^{N+,N+}_{00}(q)&=&Z_{00}(q)=\chi_+^*(q),\\ \non
Z^{N+,N-}_{00}(q)&=&\chi_-^*(q), \\ \non
Z^{N+,N}_{0m}(q)&=&\chi_m(q), \qquad 1\le m\le k-1 \\ \non
Z^{N+,N+}_{0k}(q)&=&Z^{N+,N-}_{0k}(q)= \chi_k^*(q), \\ \non
Z^{N+,D+}_{00}(q)&=&Z^{N+,D+}_{0k}(q) = \chi_+(q), \\ \non
Z^{N+,D-}_{00}(q)&=&Z^{N+,D-}_{0k}(q)= \chi_-(q).
\label{partom}
\feqr
\par Of course there is no a priori guarantee that all other partition functions are going to be integer
combinations of characters, but there is not much choice left for the states. In fact the full set of
partition functions, each of which is an integer sum of characters is contained in Appendix B.
Note that we were forced (by the twisted sector) to include Dirichlet boundary states at the orbifold
fixed points which correspond to $m=0$, $m=k$, while in the case of the torus models the space of boundary
states contained pure Newmann states.  This is because the orbifold fixed points correspond to the possible eigenvalues
of the position operator in the expansion of the free field.

\section{Module Decomposition}
In this section we study the extended algebra module decomposition of the space of states for the $c=1$ $Z_2$ orbifold
with $R^2=1/2k$.
\newline 
\newline
{\it The vacuum character:} 
The first question is what the extended symmetry algebra is.  One way to get this algebra is to study the
vacuum character.  We are going to find which subspace $H_0$ of the known overall space of states $H^U$ 
 gives
\beqr
Tr_{H_0}(q^{L_0-1/24}q^{\bar{L}_0-1/24})&=&\chi_+^*(q)\chi_+^*(q) \\ \non
&=&\left( \sum_{n=1}^{\infty }\frac{q^{kn^2}}{\eta (q)}+
\sum_{n=0}^{\infty }\frac{q^{(2n)^2}-q^{(2n+1)^2}}{\eta (q)} \right)
\left( \sum_{n=1}^{\infty }\frac{q^{kn^2}}{\eta (q)}+
\sum_{n=0}^{\infty }\frac{q^{(2n)^2}-q^{(2n+1)^2}}{\eta (q)} \right).
\label{hilb0a}
\feqr
Next we are going to determine a set of operators that generates $H_0$ from the vacuum state. The algebra of these 
operators is the extended symmetry algebra. 
\par To this end we need the following lemma \cite{ginsp}:
\begin{lemma}
\beqr
q^{-1/12}Tr_{H^U_{mn}}(q^{L_0}q^{\bar{L}_0})=\frac{1}{\eta^2(q)}q^{\frac{k}{4}(m+\frac{n}{k})^2}
q^{\frac{k}{4}(m-\frac{n}{k})^2}
+\frac{1}{2}\frac{q^{-1/12}\delta_{m,0}\delta_{n,0}}{\prod_{n=1}^{\infty }(1+q^n)^2}.
\label{trlemma}
\feqr
\end{lemma}

\par The traces that can contribute to the vacuum character are the ones for which $m=\rho $ and $n=k\rho '$ where
$\rho -\rho '=0mod2$.  Otherwise we get powers of q that correspond to noninteger conformal dimension. 
If $\rho -\rho '=2s'$ and $\rho +\rho '=2s$ then we get that the following traces can
contribute
\beqr
q^{-1/12}Tr_{H^U_{s+s',k(s-s')}}(q^{L_0}q^{\bar{L}_0})=\frac{1}{\eta^2(q)}q^{ks^2}q^{ks'^2}
+\frac{1}{2}\frac{q^{-1/12}\delta_{s,0}\delta_{s',0}}{\prod_{n=1}^{\infty }(1+q^n)^2}
\label{trlemma1}
\feqr
where $s,s'\in Z$.  Of course we really need half of these spaces because there is also the product
$\chi_-^*(q)\chi_-^*(q)$ that should be generated by traces of the same $H^U_{mn}$.  Thus we define
the following subspaces of $H^U_{s+s',k(s-s')}$, $H^U_{s,ks}$, $H^U_{s',-ks'}$ (with $s,s'>0$), and of $H^U_{0,0}$:
\beqr
\label{huhilb1}
H^{Uv}_{s+s',k(s-s')}&=&\{\alpha_{-n_1}\cdots \alpha_{-n_{2j}}\bar{\alpha}_{-n_{2j+1}}\cdots \bar{\alpha}_{-n_{2l}}
(|s+s',k(s-s')> \\ \non
&+&|-(s+s'),-k(s-s')>+|s-s',k(s+s')>+|-(s-s'),-k(s+s')>)\} \\ \non
&\oplus &\{\alpha_{-n_1}\cdots \alpha_{-n_{2j-1}}\bar{\alpha}_{-n_{2j}}\cdots \bar{\alpha}_{-n_{2l}}
(|s+s',k(s-s')> \\ \non
&+&|-(s+s'),-k(s-s')>-|s-s',k(s+s')>-|-(s-s'),-k(s+s')>)\} \\ \non
&\oplus &\{\alpha_{-n_1}\cdots \alpha_{-n_{2j-1}}\bar{\alpha}_{-n_{2j}}\cdots \bar{\alpha}_{-n_{2l+1}}
(|s+s',k(s-s')>\\ \non
&-&|-(s+s'),-k(s-s')>+|s-s',k(s+s')>-|-(s-s'),-k(s+s')>)\} \\ \non
&\oplus &\{\alpha_{-n_1}\cdots \alpha_{-n_{2j}}\bar{\alpha}_{-n_{2j+1}}\cdots \bar{\alpha}_{-n_{2l+1}}
(|s+s',k(s-s')> \\ \non
&-&|-(s+s'),-k(s-s')>-|s-s',k(s+s')>+|-(s-s'),-k(s+s')>)\},
\\ \non
H^{Uv}_{s,ks}&=&\{\alpha_{-n_1}\cdots \alpha_{-n_{2j}}\bar{\alpha}_{-n_{2j+1}}\cdots \bar{\alpha}_{-n_{2l}}
(|s,ks>+|-s,-ks>)\} \\ \non
&\oplus &\{\alpha_{-n_1}\cdots \alpha_{-n_{2j-1}}\bar{\alpha}_{-n_{2j}}\cdots \bar{\alpha}_{-n_{2l+1}}(|s,ks>-|-s,-ks>)\},
\\ \non
H^{Uv}_{s',-ks'}&=&\{\alpha_{-n_1}\cdots \alpha_{-n_{2j}}\bar{\alpha}_{-n_{2j+1}}\cdots \bar{\alpha}_{-n_{2l}}
(|s',-ks'>+|-s',ks'>)\} \\ \non
&\oplus &\{\alpha_{-n_1}\cdots \alpha_{-n_{2j}}\bar{\alpha}_{-n_{2j+1}}\cdots \bar{\alpha}_{-n_{2l+1}}(|s',-ks'>-|-s',ks'>)\},
\\ \non
H^{Uv}_{0,0}&=&\{\alpha_{-n_1}\cdots \alpha_{-n_{2j}}\bar{\alpha}_{-n_{2j+1}}\cdots \bar{\alpha}_{-n_{2l}}|0,0>\}.
\feqr
The traces over the subspaces \rf{huhilb1} are summarized in the following lemma:
\begin{lemma}
\beqr
q^{-1/12}Tr_{H^{Uv}_{s+s',k(s-s')}}(q^{L_0}q^{\bar{L}_0})&=&\frac{1}{\eta^2(q)}q^{ks^2}q^{ks'^2}, \qquad s,s'>0 \\ \non
q^{-1/12}Tr_{H^{Uv}_{s,ks}}(q^{L_0}q^{\bar{L}_0})&=&\frac{1}{\eta^2(q)}q^{ks^2}\left( \sum_{n\ge 0}q^{n^2}(-1)^n
\right), \qquad s>0 \\ \non
q^{-1/12}Tr_{H^{Uv}_{s',-ks'}}(q^{L_0}q^{\bar{L}_0})&=&\frac{1}{\eta^2(q)}\left(\sum_{n\ge 0}q^{n^2}(-1)^n\right) q^{ks'^2},
\qquad s'>0 \\ \non
q^{-1/12}Tr_{H^{Uv}_{0,0}}(q^{L_0}q^{\bar{L}_0})&=&\frac{1}{\eta^2(q)}\left( \sum_{n\ge 0}q^{n^2}(-1)^n\right)
\left( \sum_{n\ge 0}q^{n^2}(-1)^n \right).
\label{traces2}
\feqr
\end{lemma}
For the proof of this lemma see Appendix C.
\par Now we are in a position to define our space $H_0$:
\beqr
H_0=\sum_{s,s'>0}H^{Uv}_{s+s',k(s-s')}\oplus \sum_{s>0}H^{Uv}_{s,ks}\oplus \sum_{s'>0}H^{Uv}_{s',-ks'}\oplus
H^{Uv}_{0,0}.
\label{vachilb1}
\feqr
The trace over $H_0$ is
\beqr
q^{-1/12}Tr_{H_0}(q^{L_0}q^{\bar{L}_0})&=&\frac{1}{\eta^2(q)}\Biggl[\left(\sum_{s>0}q^{ks^2}\right)
\left( \sum_{s'>0}q^{ks'^2}\right) +
\left( \sum_{s>0}q^{ks^2}\right) \left( \sum_{n\ge 0}q^{n^2}(-1)^n \right) \\ \non
&+&\left(\sum_{n\ge 0}q^{n^2}(-1)^n\right) \left( \sum_{s'>0}q^{ks'^2}\right)+
\left( \sum_{n\ge 0}q^{n^2}(-1)^n\right)
\left( \sum_{n\ge 0}q^{n^2}(-1)^n \right)\Biggr] \\ \non
&=&\chi_+^*(q)\chi_+^*(q).
\label{vactr1}
\feqr

\par Having identified the vacuum module, it is interesting to try to read what the symmetry algebra is.  Recall that
in the unorbifolded theory we had three generators of the algebra \cite{vafa}, 
$J^0(z)=\frac{1}{\sqrt{2k}}i\partial \phi (z)$ and
$J^{\pm }(z)=e^{\pm i\sqrt{2k}\phi (z)}$.  Now we have to select elements of the algebra generated by the modes that
remain invariant under the $Z_2$ symmetry.  Such elements are the following:
\beqr
S=\{ a_{n_i}a_{n_j}, J^+_k+J^-_k, a_{n}(J^+_k-J^-_k),\bar{a}_{n_i}\bar{a}_{n_j}, \bar{J}^+_k+\bar{J}^-_k,
\bar{a}_{n}(\bar{J}^+_k-\bar{J}^-_k)\}.
\label{symmalg}
\feqr
Of course these are not the only ones.  Now we have the following proposition:
\begin{proposition}
Starting from a vacuum $|0,0>$, S generates precisely the module $H_0$.
\end{proposition}
For the proof of this proposition see Appendix D.
So the symmetry algebra of the theory is the algebra generated by the elements of S.
\newline
\newline
{\it The character $\chi_-^*(q)$: } 
The space $H_1$ that generates the product $\chi_-^*(q)\chi_-^*(q)$ can be
constructed similarly. Define
\beqr
H^{Ub}_{s+s',k(s-s')}&=&\{\alpha_{-n_1}\cdots \alpha_{-n_{2j-1}}\bar{\alpha}_{-n_{2j}}\cdots \bar{\alpha}_{-n_{2l}}
(|s+s',k(s-s')> \\ \non
&+&|-(s+s'),-k(s-s')>+|s-s',k(s+s')>+|-(s-s'),-k(s+s')>)\} \\ \non
&\oplus &\{\alpha_{-n_1}\cdots \alpha_{-n_{2j}}\bar{\alpha}_{-n_{2j+1}}\cdots \bar{\alpha}_{-n_{2l}}
(|s+s',k(s-s')> \\ \non
&+&|-(s+s'),-k(s-s')>-|s-s',k(s+s')>-|-(s-s'),-k(s+s')>)\} \\ \non
&\oplus &\{\alpha_{-n_1}\cdots \alpha_{-n_{2j}}\bar{\alpha}_{-n_{2j+1}}\cdots \bar{\alpha}_{-n_{2l+1}}
(|s+s',k(s-s')>\\ \non
&-&|-(s+s'),-k(s-s')>+|s-s',k(s+s')>-|-(s-s'),-k(s+s')>)\} \\ \non
&\oplus &\{\alpha_{-n_1}\cdots \alpha_{-n_{2j-1}}\bar{\alpha}_{-n_{2j}}\cdots \bar{\alpha}_{-n_{2l+1}}
(|s+s',k(s-s')> \\ \non
&-&|-(s+s'),-k(s-s')>-|s-s',k(s+s')>+|-(s-s'),-k(s+s')>)\},
\\ \non
H^{Ub}_{s,ks}&=&\{\alpha_{-n_1}\cdots \alpha_{-n_{2j-1}}\bar{\alpha}_{-n_{2j}}\cdots \bar{\alpha}_{-n_{2l}}
(|s,ks>+|-s,-ks>)\} \\ \non
&\oplus &\{\alpha_{-n_1}\cdots \alpha_{-n_{2j}}\bar{\alpha}_{-n_{2j+1}}\cdots \bar{\alpha}_{-n_{2l+1}}(|s,ks>-|-s,-ks>)\},
\\ \non
H^{Ub}_{s',-ks'}&=&\{\alpha_{-n_1}\cdots \alpha_{-n_{2j-1}}\bar{\alpha}_{-n_{2j}}\cdots \bar{\alpha}_{-n_{2l}}
(|s',-ks'>+|-s',ks'>)\} \\ \non
&\oplus &\{\alpha_{-n_1}\cdots \alpha_{-n_{2j-1}}\bar{\alpha}_{-n_{2j}}\cdots \bar{\alpha}_{-n_{2l+1}}(|s',-ks'>-|-s',ks'>)\},
\\ \non
H^{Ub}_{0,0}&=&\{\alpha_{-n_1}\cdots \alpha_{-n_{2j-1}}\bar{\alpha}_{-n_{2j}}\cdots \bar{\alpha}_{-n_{2l}}|0,0>\}.
\label{huhilb2}
\feqr
Supplied with these definitions we have
\beqr
H_1=\sum_{s,s'>0}H^{Ub}_{s+s',k(s-s')}\oplus \sum_{s>0}H^{Ub}_{s,ks}\oplus \sum_{s'>0}H^{Ub}_{s',-ks'}\oplus
H^{Ub}_{0,0}.
\label{bhilb1}
\feqr
The associated traces are computed in the following lemma:
\begin{lemma}
\beqr
q^{-1/12}Tr_{H^{Ub}_{s+s',k(s-s')}}(q^{L_0}q^{\bar{L}_0})&=&\frac{1}{\eta^2(q)}q^{ks^2}q^{ks'^2}, \qquad s,s'>0 \\ \non
q^{-1/12}Tr_{H^{Ub}_{s,ks}}(q^{L_0}q^{\bar{L}_0})&=&\frac{1}{\eta^2(q)}q^{ks^2}\left( \sum_{n\ge 1}q^{n^2}(-1)^{n-1}
\right), \qquad s>0 \\ \non
q^{-1/12}Tr_{H^{Ub}_{s',-ks'}}(q^{L_0}q^{\bar{L}_0})&=&\frac{1}{\eta^2(q)}\left(\sum_{n\ge 1}q^{n^2}(-1)^{n-1}\right)
q^{ks'^2},
\qquad s'>0 \\ \non
q^{-1/12}Tr_{H^{Ub}_{0,0}}(q^{L_0}q^{\bar{L}_0})&=&\frac{1}{\eta^2(q)}\left( \sum_{n\ge 1}q^{n^2}(-1)^{n-1}\right)
\left( \sum_{n\ge 1}q^{n^2}(-1)^{n-1} \right).
\label{traces3}
\feqr
\end{lemma}
The proof of this lemma is similar to the proof of lemma 3.  
So finally we have for the trace over $H_1$
\beqr
q^{-1/12}Tr_{H_1}(q^{L_0}q^{\bar{L}_0})&=&\frac{1}{\eta^2(q)}\Biggl[\left(\sum_{s>0}q^{ks^2}\right)
\left( \sum_{s'>0}q^{ks'^2}\right) +
\left( \sum_{s>0}q^{ks^2}\right) \left( \sum_{n\ge 1}q^{n^2}(-1)^{n-1} \right) \\ \non
&+&\left(\sum_{n\ge 1}q^{n^2}(-1)^{n-1}\right) \left( \sum_{s'>0}q^{ks'^2}\right)+
\left( \sum_{n\ge 1}q^{n^2}(-1)^{n-1}\right)
\left( \sum_{n\ge 1}q^{n^2}(-1)^{n-1} \right)\Biggr] \\ \non
&=&\chi_-^*(q)\chi_-^*(q).
\label{vactr2}
\feqr
Now we have the following proposition:
\begin{proposition}
Starting from the state $a_{-1}\bar{a}_{-1}|0,0>$, S generates precisely the module $H_1$.
\end{proposition}
For the proof of this proposition see Appendix D.
\par Next we ask what the direct sum $H_0\oplus H_1$ is.  To find this we need the identities
\beqr
H^{Uv}_{s+s',k(s-s')}\oplus H^{Ub}_{s+s',k(s-s')} &= &H^U_{s+s',k(s-s')}\oplus H^U_{s-s',k(s+s')}, \qquad s,s'> 0 \\ \non
H^{Uv}_{s,ks}\oplus H^{Ub}_{s,ks}&=&H^U_{s,ks}, \qquad s>0 \\ \non
H^{Uv}_{s',-ks'}\oplus H^{Ub}_{s',-ks'}&=&H^U_{s',-ks'}, \qquad s'>0 \\ \non
H^{Uv}_{0,0}\oplus H^{Ub}_{0,0}&=&H^U_{0,0}
\label{hilbid1}
\feqr
where it is understood that $H^U_{s-s',k(s+s')}=H^U_{s'-s,-k(s+s')}$ if $s-s'<0$.
Summing over $s,s'\ge 0$ we get
\beqr
H_0\oplus H_1&=&\oplus_{s,s'> 0}\left(H^U_{s+s',k(s-s')}\oplus H^U_{s-s',k(s+s')}\right)\oplus_{s>0}H^U_{s,ks}
\oplus_{s'>0}H^U_{s',-ks'}\oplus H^U_{0,0} \\ \non
&=&\oplus_{\stackrel{s+s'> 0}{s,s'\in Z}}H^U_{s+s',k(s-s')}\oplus_{s\ge 0}H_{0,2ks}
\label{hilbsum1}
\feqr
where the condition $s+s'>0$ corresponds to the condition $m>0$ (if $m=0$, $n\ge 0$) in the original Hilbert space.
\newline
\newline
{\it The characters $\chi_l(q)$:} 
The next question is what module $H_{l^2/4k}$  gives rise to the product $\chi_l(q)\chi_l(q)$ for $0<l<k$.  The
spaces that can contribute to this product have to produce the correct powers of q in the trace.  This requires
\beqr
\label{condi2}
\frac{1}{2}(\frac{m}{2R}+nR)^2=\frac{(mk+n)^2}{4k}=\frac{l^2}{4k}modZ.
\feqr
Equation \rf{hilbsum1} is satisfied provided that $m=\rho$ and $n=l+k\rho ' $ where $\rho -\rho '=0mod2$.
Defining again $2s=\rho +\rho '$ and $2s'=\rho -\rho '$ we get
that only the modules $H^U_{s+s',k(s-s')+l}$ can contribute.
It is natural to guess that $H_{l^2/4k}=\oplus_{s,s'\in Z}H^U_{s+s',k(s-s')+l}$.
The modules $H^U_{s+s',k(s-s')+l}$ with $s+s'<0$ make sense through  
the identity $H^U_{s+s',k(s-s')+l}=H^U_{-(s+s'),-k(s-s'+2)+2k-l}$.  So now we have 
\beqr
H_{l^2/4k}&=&\oplus_{\stackrel{s+s'>0}{if \, s+s'=0 \, then \, s\ge 0}}H^U_{s+s',k(s-s')+l}
\oplus_{\stackrel{s+s'<0}{if \, s+s'=0\, then \, s< 0}}H^U_{-(s+s'),-k(s-s'+2)+2k-l} \\ \non 
&=& \oplus_{\stackrel{s+s'>0}{if \, s+s'=0 \, then \, s\ge 0}}H^U_{s+s',k(s-s')+l}
\oplus_{\stackrel{s+s'>0}{if \, s+s'=0\, then \, s\ge 0}}H^U_{(s+s'),k(s-s')+2k-l}.
\label{hilbsum2}
\feqr
Summing over l for $1\le l\le k-1$ we get 
\beqr
\oplus_{l=1}^{k-1}H_{l^2/4k}&=& \oplus_{l=1}^{k-1}\oplus_{\stackrel{s+s'>0}{if \, s+s'=0 \, then \, s\ge 0}}
H^U_{s+s',k(s-s')+l}
\oplus_{\stackrel{s+s'>0}{if \, s+s'=0\, then \, s\ge 0}}H^U_{(s+s'),k(s-s')+k+l} \\ \non 
&=&\oplus_{l=1}^{k-1}\oplus_{\stackrel{s+s'>0}{if \, s+s'=0 \, then \, s\ge 0}}H^U_{s+s',k(s-s')+l}
\oplus_{\stackrel{s+s'>0}{if \, s+s'=0\, then \, s\ge 0}}H^U_{(s+s'),k(s-s'+1)+l} \\ \non 
&=&\oplus_{l=1}^{k-1}\oplus_{\stackrel{\rho >0,\rho '}{if \, \rho =0\, then \, \rho '\ge 0}}H^U_{\rho ,k\rho '+l}.
\label{hilbsum3}
\feqr 
Now we have the following lemma:
\begin{lemma}
\beqr 
q^{-1/12}Tr_{H_{l^2/4k}}(q^{L_0}q^{\bar{L}_0})=
q^{-1/12}\sum_{s,s'\in Z}Tr_{H^U_{s+s',k(s-s')+l}}(q^{L_0}q^{\bar{L}_0}) =
\chi_l(q)\chi_l(q).
\label{trace4}
\feqr
\end{lemma} 
The proof of this lemma depends on the trace
\beqr 
q^{-1/12}Tr_{H^U_{s+s',k(s-s')+l}}(q^{L_0}q^{\bar{L}_0})&=&\frac{1}{\eta^2(q)}q^{k(s+l/2k)^2}q^{k(s'-l/2k)^2}, 
\qquad s,s'\in Z 
\label{trace5}
\feqr
which comes from lemma 2. Now we have the following proposition:
\begin{proposition}
Starting from the state $|0,l>+|0,-l>$, S generates precisely the module $H_{l^2/4k}$.
\end{proposition}
\noi
{\it The character $\chi^*_k(q)$:}
Now we need the modules $H^{v}_{k/4}$, $H^{b}_{k/4}$ that give rise to the product $2\chi^*_k(q)\chi^*_k(q)$. 
The modules that can contribute to 
this product are $H^U_{\rho ,k\rho '}$ where $\rho -\rho '=1mod2$.  So we introduce the following spaces:
\scriptsize
\beqr 
H^{Uv}_{s+s'+1,k(s-s')}&=& \{\alpha_{-n_1}\cdots \alpha_{-n_{2j}}\bar{\alpha}_{-n_{2j+1}}\cdots \bar{\alpha}_{-n_{2l}}
(|s+s'+1,k(s-s')> \\ \non
&+& |-(s+s'+1),-k(s-s')>+|s-s',k(s+s'+1)>+|-(s-s'),-k(s+s'+1)>)\} \\ \non
&\oplus &\{\alpha_{-n_1}\cdots \alpha_{-n_{2j-1}}\bar{\alpha}_{-n_{2j}}\cdots \bar{\alpha}_{-n_{2l}}
(|s+s'+1,k(s-s')> \\ \non
&+&|-(s+s'+1),-k(s-s')>-|s-s',k(s+s'+1)>-|-(s-s'),-k(s+s'+1)>)\} \\ \non
&\oplus &\{\alpha_{-n_1}\cdots \alpha_{-n_{2j-1}}\bar{\alpha}_{-n_{2j}}\cdots \bar{\alpha}_{-n_{2l+1}}
(|s+s'+1,k(s-s')>\\ \non
&-&|-(s+s'+1),-k(s-s')>+|s-s',k(s+s'+1)>-|-(s-s'),-k(s+s'+1)>)\} \\ \non
&\oplus &\{\alpha_{-n_1}\cdots \alpha_{-n_{2j}}\bar{\alpha}_{-n_{2j+1}}\cdots \bar{\alpha}_{-n_{2l+1}}
(|s+s'+1,k(s-s')> \\ \non
&-&|-(s+s'+1),-k(s-s')>-|s-s',k(s+s'+1)>+|-(s-s'),-k(s+s'+1)>)\},
\\ \non 
H^{Ub}_{s+s'+1,k(s-s')}&=&\{\alpha_{-n_1}\cdots \alpha_{-n_{2j-1}}\bar{\alpha}_{-n_{2j}}\cdots \bar{\alpha}_{-n_{2l}}
(|s+s'+1,k(s-s')> \\ \non
&+&|-(s+s'+1),-k(s-s')>+|s-s',k(s+s'+1)>+|-(s-s'),-k(s+s'+1)>)\} \\ \non
&\oplus &\{\alpha_{-n_1}\cdots \alpha_{-n_{2j}}\bar{\alpha}_{-n_{2j+1}}\cdots \bar{\alpha}_{-n_{2l}}
(|s+s'+1,k(s-s')> \\ \non
&+&|-(s+s'+1),-k(s-s')>-|s-s',k(s+s'+1)>-|-(s-s'),-k(s+s'+1)>)\} \\ \non
&\oplus &\{\alpha_{-n_1}\cdots \alpha_{-n_{2j}}\bar{\alpha}_{-n_{2j+1}}\cdots \bar{\alpha}_{-n_{2l+1}}
(|s+s'+1,k(s-s')>\\ \non
&-&|-(s+s'+1),-k(s-s')>+|s-s',k(s+s'+1)>-|-(s-s'),-k(s+s'+1)>)\} \\ \non
&\oplus &\{\alpha_{-n_1}\cdots \alpha_{-n_{2j-1}}\bar{\alpha}_{-n_{2j}}\cdots \bar{\alpha}_{-n_{2l+1}}
(|s+s'+1,k(s-s')> \\ \non
&-&|-(s+s'+1),-k(s-s')>-|s-s',k(s+s'+1)>+|-(s-s'),-k(s+s'+1)>)\}.
\label{hilbsum6}
\feqr
\normalsize
These definitions lead to 
\beqr 
H^v_{k/4}&=&\oplus_{s,s'\ge 0}H^{Uv}_{s+s'+1,k(s-s')}, \\ \non 
H^b_{k/4}&=&\oplus_{s,s'\ge 0}H^{Ub}_{s+s'+1,k(s-s')}.
\label{hilbsum7}
\feqr 
The traces over these spaces are computed in the following lemma:
\begin{lemma}
\beqr 
q^{-1/12}Tr_{H^v_{k/4}}(q^{L_0}q^{\bar{L}_0})=q^{-1/12}Tr_{H^b_{k/4}}(q^{L_0}q^{\bar{L}_0})=
\chi^*_k(q)\chi^*_k(q).
\label{trace8}
\feqr
\end{lemma}
Note that 
\beqr
\bigoplus_{s,s'\ge 0}H^{Uv}_{s+s'+1,k(s-s')}\oplus H^{Ub}_{s+s'+1,k(s-s')}=\bigoplus_{\stackrel{s+s'\ge 0}{s,s'\in Z}}
H^U_{s+s'+1,k(s-s')}\oplus_{s\ge 0}H^U_{0,k(2s+1)}.
\label{hilbsum8}
\feqr 
Furthermore we have the additional proposition:
\begin{proposition}
Starting from the states $|1,0> +|-1,0>+|0,k>+|0,-k>$ and $|1,0> +|-1,0>-|0,k>-|0,-k>$ , S generates precisely 
the modules $H^v_{k/4}$ and $H^b_{k/4}$.
\end{proposition}

\par So we have identified completely the extended symmetry algebra to be the algebra of products of elements of S. 
The generating functions 
that give rise to the elements of S are nonlocal as expected, since the orbifold symmetry is nonlocal, and are:
\beqr 
J^0(z)J^0(w), \quad J^+(z)+J^-(z), \quad J^0(z)(J^+(w)-J^-(w)) \quad and \; c. \, c.
\label{alg1}
\feqr 
Again this algebra splits into a chiral and an antichiral part. Since 
\beqr
H_0 \oplus H_1 \oplus_{l=1}^{k-1}H_{l^2/4k}\oplus H^v_{k/4}\oplus H^b_{k/4}=H^U
\label{ovsp}
\feqr
we come to the conclusion that the periodic part of the Hilbert space of the theory splits into a sum of 
modules of the extended symmetry algebra, 
each module giving rise to a product of extended characters that appear in the partition function of the theory.  
\newline
\newline
{\it The characters $\chi_+(q)$, $\chi_-(q)$:}
Now consider the antiperiodic sector.  Upon changing the boundary conditions the only elements of S that 
survive are $a_{n_1}a_{n_2}$ and $\bar{a}_{n_1}\bar{a}_{n_2}$ where now $n_1$ and $n_2$ become half integers. 
The currents dissapear since in the antiperiodic sector there is no charge changing operator.  
Now we define the modules  
\beqr
H^{Te}_{0,\pi R}&=&\alpha_{-n_1}\cdots \alpha_{-n_{2j}}\bar{\alpha}_{-n_{2j+1}}\cdots \bar{\alpha}_{-n_{2l}}
|1/16,1/16>_{0,\pi R},\\ \non 
H^{To}_{0,\pi R}&=&\alpha_{-n_1}\cdots \alpha_{-n_{2j-1}}\bar{\alpha}_{-n_{2j}}\cdots \bar{\alpha}_{-n_{2l}}
|1/16,1/16>_{0,\pi R}.
\label{thilb1}
\feqr 
We have the following lemma:
\begin{lemma}
\beqr 
q^{-1/12}Tr_{H^{Te}_{0,\pi R}}(q^{L_0}q^{\bar{L}_0})&=&\chi_+(q)\chi_+(q)\\ \non 
q^{-1/12}Tr_{H^{To}_{0,\pi R}}(q^{L_0}q^{\bar{L}_0})&=&\chi_-(q)\chi_-(q)
\label{trace10}
\feqr
\end{lemma}
Finally we have the proposition:
\begin{proposition}
Starting from the states $|1/16,1/16>_{0,\pi R}$ and $a_{-1/2}\bar{a}_{-1/2}|1/16,1/16>$ , the surviving elements of S generate precisely the modules $H^{Te}_{0,\pi R}$ and $H^{To}_{0,\pi R}$.
\end{proposition}

\par It is worth mentioning that the new definitions of modules simply give another decomposition of the 
original Hilbert space into modules of the orbifold algebra.  So the direct sum of the modules that give rise to
the products of characters gives the original Hilbert space.

\section{Conclusions}

We studied the boundary state structure of the $Z_2$ orbifold models with $R^2=1/2k$.  We found a complete set of
boundary states that satisfies both the Virasoro Ishibashi condition and the Cardy condition.  It is interesting to
emphasize that in the case of the $R^2=1/2k$ torus models all the boundary states were Newmann states while in the 
orbifold case
we were forced to introduce four Dirichlet states, one pair corresponding to each orbifold fixed point.  Another point
of interest is that the decomposition of the extended algebra characters into a sum of Virasoro characters changes
nature in the case $k=l^2$.  The models with this k are identified to be the $D_l$ dihedral group orbifolds
of the $SU(2)\times SU(2)$ level one theory, which corresponds to the $k=1$ ($R=1/\sqrt{2}$) torus model. Note that the 
T-duality $R\rightarrow 1/2R$ just exchange Newmann and Dirichlet states preserving the essential
structure of the theory.
\par Finally, the space of states of the theory admits a decomposition into highest weight modules of the 
extended symmetry algebra.  It is worth noting that two of the generating functions of our set of generators are 
nonlocal fields (they are bilocal), but they respect chirality.  This is of course expected, since the $Z_2$ 
symmmetry is a global symmetry of the theory.

\addcontentsline{toc}{subsection}{Appendix A }
\section*{Appendix A }
\renewcommand{\theequation}{A.\arabic{equation}}
\setcounter{equation}{0}

The elliptic theta functions and the Dedekind eta function are:
\beqr
\theta_1(w,q)&=& i \sum_{n=-\infty}^{\infty}(-1)^n q^{\frac{1}{2}(n-\frac{1}{2})^2}w^{n-\frac{1}{2}}, \\ \non
\theta_2(w,q) &=& \sum_{n=-\infty}^{\infty}q^{\frac{1}{2}(n-\frac{1}{2})^2}w^{n-\frac{1}{2}}, \\ \non
\theta_3(w,q)&=&  \sum_{n=-\infty}^{\infty}q^{\frac{1}{2}n^2}w^n,  \\ \non
\theta_4(w,q)&=&  \sum_{n=-\infty}^{\infty}(-1)^n q^{\frac{1}{2}n^2}w^n,  \\ \non
\eta(q)&=& q^{\frac{1}{24}}\prod_{n=1}^{\infty}(1-q^n).
\label{theta}
\feqr

{\it Proof of Lemma 1:} The characters of the Virasoro algebra with $c=1$ are
\beqr
\chi_h^{Vir}(q)&=&\frac{q^h}{\eta (q)}, \qquad h\ne n^2/4 \\ \non
\chi_{n^2}^{Vir}(q)&=&\frac{q^{n^2}-q^{(n+1)^2}}{\eta(q)}, \\ \non
\chi_{(n+1/2)^2}^{Vir}(q)&=&\frac{q^{(n+1/2)^2}-q^{(n+3/2)^2}}{\eta(q)}.
\label{virchar}
\feqr
First we consider the components in the case $k\ne l^2$.  These are
\beqr
\chi^*_+(q)&=&\sum_{n=1}^{\infty }\frac{q^{kn^2}}{\eta (q)}+\sum_{n=0}^{\infty }\frac{q^{(2n)^2}-q^{(2n+1)^2}}{\eta (q)},
\\ \non
\chi^*_-(q)&=&\sum_{n=1}^{\infty }\frac{q^{kn^2}}{\eta (q)}+\sum_{n=0}^{\infty }\frac{q^{(2n+1)^2}-q^{(2n+2)^2}}{\eta (q)},
\\ \non
\chi^*_k(q)&=&\sum_{n=0}^{\infty }\frac{q^{k(n+1/2)^2}}{\eta(q)}, \\ \non
\chi_s(q)&=&\sum_{n\in Z}\frac{q^{k(n+\frac{s}{2k})^2}}{\eta(q)}, \qquad 1\le s \le k-1 \\ \non
\chi_+(q)&=&\frac{1}{\sqrt{2}}(\chi_a(q)+\chi_b(q))=\sum_{n\in Z}\frac{q^{4(n+1/8)^2}}{\eta (q)}, \\ \non
\chi_-(q)&=&\frac{1}{\sqrt{2}}(\chi_a(q)-\chi_b(q))=\sum_{n\in Z}\frac{q^{4(n+3/8)^2}}{\eta (q)}
\label{virsplit}
\feqr
where the splitting of the components into a sum of Virasoro characters is exhibited.  
This splitting is no longer valid
for all components
if $k=l^2$.  This is justified by the fact that some powers of q are of the form $q^{n^2/4}$ and $q^{n^2/4}/\eta (q)$ is
not a Virasoro character.
The characters that contain such powers are $\chi^*_\pm (q)$, $\chi^*_k(q)$ and $\chi_s(q)$ for $s=pl$, $1\le p \le l-1$.
To deduce the splitting in this case we consider the sum
\beqr
&& \sum_{n\ge 0}\frac{q^{k(n+s/2k)^2}}{\eta (q)}= \sum_{n\ge 0}\frac{q^{\frac{(2nl+p)^2}{4}}}{\eta (q)} \\ \non
&=&\frac{1}{\eta (q)}\Bigl( (q^{p^2/4}-q^{(p+2)^2/4})+(q^{(p+2)^2/4}-q^{(p+4)^2/4})
+\cdots +(q^{(p+2l-2)^2/4}-q^{(p+2l)^2/4}) \\ \non
&+&2(q^{(p+2l)^2/4}-q^{(p+2l+2)^2/4})+\cdots + 2(q^{(p+4l-2)^2/4}-q^{(p+4l)^2/4}) \\ \non
&+&3(q^{(p+4l)^2/4}-q^{(p+4l+2)^2/4})+\cdots + 3(q^{(p+6l-2)^2/4}-q^{(p+6l)^2/4}) \\ \non
&+& \cdots \Bigr) \\ \non
&=& \sum_{n=0}^{\infty } \sum_{m=0}^{l-1}(n+1)\chi^{Vir}_{(p/2+nl+m)^2}(q).
\label{idvir}
\feqr
Note that
\beqr
\chi_{pl}(q)=\sum_{n\ge 0}\frac{q^{(2nl+p)^2/4}}{\eta (q)}+\sum_{n\ge 0}\frac{q^{(2nl+2l-p)^2/4}}{\eta (q)}.
\label{split1}
\feqr
Now the splitting of the problematic components takes the form
\beqr
\chi^*_+(q)&=& \sum_{n=1}^{\infty } \sum_{m=0}^{l-1}n\chi^{Vir}_{(nl+m)^2}(q)+
\sum_{n=0}^{\infty }\chi^{Vir}_{(2n)^2}(q), \\ \non
\chi^*_-(q)&=& \sum_{n=1}^{\infty } \sum_{m=0}^{l-1}n\chi^{Vir}_{(nl+m)^2}(q)+
\sum_{n=0}^{\infty }\chi^{Vir}_{(2n+1)^2}(q), \\ \non
\chi^*_k(q)&=& \sum_{n=0}^{\infty } \sum_{m=0}^{l-1}(n+1)\chi^{Vir}_{(l/2+nl+m)^2}(q), \\ \non
\chi_s(q) &=& \sum_{n=0}^{\infty } \sum_{m=0}^{l-1}(n+1)\chi^{Vir}_{(p/2+nl+m)^2}(q) +
\sum_{n=1}^{\infty } \sum_{m=0}^{l-1}n\chi^{Vir}_{(-p/2+nl+m)^2}(q),
\label{virsplit2}
\feqr
while the rest of the components preserves its splitting.

\addcontentsline{toc}{subsection}{Appendix B }
\section*{Appendix B }
\renewcommand{\theequation}{B.\arabic{equation}}
\setcounter{equation}{0}

The partition functions corresponding to combinations of the boundary states that do not involve the vacuum boundary
state are:
\beqr
Z^{N-,N-}_{00}(q)&=&\chi_+^*(q), \\ \non
Z^{N-,N}_{0m}(q)&=&\chi_m(q), \\ \non
Z^{N-,N\pm }_{0k}(q)&=&\chi_k^*(q), \\ \non
Z^{N-,D\pm }_{00}(q)&=&\chi_\mp (q), \\ \non
Z^{N-,D\pm }_{0k}(q)&=&\chi_\mp (q), \\ \non
Z^{N,N}_{m_1m_2}(q)&=&2\chi_{m_1-m_2}(q), \\ \non
Z^{N,N\pm }_{mk}(q)&=&\chi_{k-m}(q), \\ \non
Z^{N,D\pm }_{m0}(q)&=&\chi_+(q)+\chi_-(q), \\ \non
Z^{N,D\pm }_{mk}(q)&=&\chi_+(q)+\chi_-(q), \\ \non
Z^{N+,N\pm }_{kk}(q)&=&\chi_{\pm}^*(q), \\ \non
Z^{N+,D\pm }_{k0}(q)&=&\chi_{\pm}(q), \\ \non
Z^{N+,D\pm }_{kk}(q)&=&\chi_{\mp}(q), \\ \non
Z^{N-,N- }_{kk}(q)&=&\chi_+^*(q), \\ \non
Z^{N-,D\pm }_{k0}(q)&=&\chi_{\mp }(q), \\ \non
Z^{N-,D\pm }_{kk}(q)&=&\chi_{\pm}(q), \\ \non
Z^{D+,D\pm }_{00}(q) &=& \left \{
\begin{array}{ll}
\chi_\pm ^*(q)+\sum_{l=1}^{k/2-1}\chi_{2l}(q)+\chi_k^*(q), & \mathrm{if \; k=even} \\
\chi_\pm ^*(q)+\sum_{l=1}^{(k-1)/2}\chi_{2l}(q), & \mathrm{if \; k=odd}
\end{array} \right. \\ \non
Z^{D+,D\pm }_{0k}(q) &=& \left \{
\begin{array}{ll}
\sum_{l=0}^{k/2-1}\chi_{2l+1}(q), & \mathrm{if \; k=even}  \\
\sum_{l=0}^{(k-1)/2 -1}\chi_{2l+1}(q)+\chi_k^*(q), & \mathrm{if \; k=odd} 
\end{array} \right. \\ \non
Z^{D-,D-}_{00}(q) &=& \left \{
\begin{array}{ll}
\chi_+^*(q)+\sum_{l=1}^{k/2-1}\chi_{2l}(q)+\chi_k^*(q), & \mathrm{if \; k=even}  \\
\chi_+^*(q)+\sum_{l=1}^{(k-1)/2}\chi_{2l}(q), & \mathrm{if \; k=odd} 
\end{array} \right. \\ \non
Z^{D-,D\pm }_{0k}(q) &=& \left \{
\begin{array}{ll}
\sum_{l=0}^{k/2-1}\chi_{2l+1}(q), & \mathrm{if \; k=even}  \\
\sum_{l=0}^{(k-1)/2 -1}\chi_{2l+1}(q)+\chi_k^*(q), & \mathrm{if \; k=odd} 
\end{array} \right. \\ \non
Z^{D+,D\pm }_{kk}(q) &=& \left \{
\begin{array}{ll}
\chi_\pm ^*(q)+\sum_{l=1}^{k/2-1}\chi_{2l}(q)+\chi_k^*(q), & \mathrm{if \; k=even}  \\
\chi_\pm ^*(q)+\sum_{l=1}^{(k-1)/2}\chi_{2l}(q), & \mathrm{if \; k=odd} 
\end{array} \right. \\ \non
Z^{D-,D-}_{kk}(q) &=& \left \{
\begin{array}{ll}
\chi_+^*(q)+\sum_{l=1}^{k/2-1}\chi_{2l}(q)+\chi_k^*(q), & \mathrm{if \; k=even}  \\ \non
\chi_+^*(q)+\sum_{l=1}^{(k-1)/2}\chi_{2l}(q), & \mathrm{if \; k=odd}. 
\end{array} \right. \\ \non
\label{partmn}
\feqr

\par These relations are just another way of writing the fusion rules of the theory, which we could have
obtained also from the S matrix through the Verlinde formula.  Of course all partition functions are positive
integer sums of characters of the extended algebra, as expected.

\addcontentsline{toc}{subsection}{Appendix C }
\section*{Appendix C }
\renewcommand{\theequation}{C.\arabic{equation}}
\setcounter{equation}{0}

\par{\it Proof of Lemma 3:} To prove the first equation we need to realize that all the states 
\scriptsize
\beqr
\pm |s+s',k(s-s')> \pm |-(s+s'),-k(s-s')>\pm |s-s',k(s+s')>\pm |-(s+s'),-k(s-s')>
\label{grl0}
\feqr
\normalsize
have the same $L_0$ and $\bar{L}_0$ eigenvalues, $ks^2$ and $ks'^2$, and so 
\beqr 
q^{-1/12}Tr_{H^{Uv}_{s+s',k(s-s')}}(q^{L_0}q^{\bar{L}_0})=q^{-1/12}Tr_{H^0_{s,s'}}(q^{L_0}q^{\bar{L}_0})
\label{aptr1}
\feqr
where 
\beqr 
H^0_{s,s'}=a_{-n_1}\cdots a_{-n_j}\bar{a}_{-n_{j+1}}\cdots \bar{a}_{-n_l}|v_{s,s'}>
\label{apthilb1}
\feqr
and $L_0|v_{s,s'}>=ks^2|v_{s,s'}>$, $\bar{L}_0|v_{s,s'}>=ks'^2|v_{s,s'}>$ . 
Now we have that 
\beqr 
q^{-1/12}Tr_{H^0_{s,s'}}(q^{L_0}q^{\bar{L}_0})=\left( q^{-1/24}Tr_{H^{0c}_{s,s'}}(q^{L_0})\right)
\left( q^{-1/24}Tr_{H^{0\bar{c}}_{s,s'}}(q^{\bar{L}_0})\right)
\label{apctr1}
\feqr
where 
\beqr 
H^{0c}_{s,s'}=a_{-n_1}\cdots a_{-n_j}|v_{s,s'}>
\label{apchilb1}
\feqr
and $H^{0\bar{c}}_{s,s'}$ is the conjugate of $H^{0c}_{s,s'}$.  
But 
\beqr 
q^{-1/24}Tr_{H^{0c}_{s,s'}}(q^{L_0})=q^{-1/24}q^{ks^2}\sum_{n=0}^{\infty } P(n)q^n =\frac{q^{ks^2}}{\eta (q)}
\label{aptr2}
\feqr
where $P(n)$ is the number of partitions of n.  So we finally have 
\beqr 
q^{-1/12}Tr_{H^0_{s,s'}}(q^{L_0}q^{\bar{L}_0})=\frac{q^{ks^2}}{\eta (q)}\frac{q^{ks'^2}}{\eta (q)}.
\label{aptr3}
\feqr 
\par To prove the remaining equations, we need to introduce the projection operators $(1+g_c)/2$, $(1-g_c)/2$, 
$(1+g_a)/2$ and $(1-g_a)/2$.  Here $g_c$ maps $a_n$ to $-a_n$ and $g_a$ maps $\bar{a}_n$ to $-\bar{a}_n$.  
The traces in the presence of the projection operators become:
\beqr 
q^{-1/24}Tr_{H^{0\bar{c}}_{s,0}}(\frac{(1+g_a)}{2}q^{\bar{L}_0})&=&
\frac{1}{2}q^{-1/24}Tr_{H^{0\bar{c}}_{s,0}}(q^{\bar{L}_0})
+\frac{1}{2}q^{-1/24}Tr_{H^{0\bar{c}}_{s,0}}(g_aq^{\bar{L}_0})\\ \non
&=&\frac{1}{2}q^{-1/24}\sum_{n\ge 0}P(n)q^n+\frac{1}{2}q^{-1/24}\prod_{n=1}^{\infty}\sum_{l\ge 0}(-1)^lq^{nl}\\ \non 
&=&\frac{1}{2\eta(q)}+\frac{1}{2}q^{-1/24}\prod_{n=1}^{\infty}\frac{1}{1+q^n}=
\frac{1}{2\eta(q)}+\frac{\sum_{n\in Z}(-1)^nq^{n^2}}{2\eta(q)}\\ \non 
&=&\sum_{n\ge 0}\frac{(-1)^nq^{n^2}}{\eta(q)}.
\label{apidtr1}
\feqr 
Similarly we get 
\beqr 
q^{-1/24}Tr_{H^{0\bar{c}}_{s,0}}(\frac{1-g_a}{2}q^{\bar{L}_0})=\sum_{n\ge 1}\frac{(-1)^{n-1}q^{n^2}}{\eta(q)}
\label{apidtr2}
\feqr
and the conjugate formulas.  

Now we have that 
\beqr 
q^{-1/12}Tr_{H^{Uv}_{s,ks}}(q^{L_0}q^{\bar{L}_0})&=&q^{-1/24}Tr_{H^{0c}_{s,0}}(q^{L_0})\cdot 
q^{-1/24}Tr_{H^{0\bar{c}}_{s,0}}(\frac{(1+g_a)}{2}q^{\bar{L}_0})\\ \non 
&=& \frac{q^{ks^2}}{\eta (q)}\sum_{n\ge 0}\frac{q^{n^2}(-1)^n}{\eta(q)},
\label{aptr4}
\feqr 
\beqr 
q^{-1/12}Tr_{H^{Uv}_{s',-ks'}}(q^{L_0}q^{\bar{L}_0})&=&q^{-1/24}Tr_{H^{0c}_{0,s'}}(\frac{(1+g_c)}{2}q^{L_0})\cdot 
q^{-1/24}Tr_{H^{0\bar{c}}_{s,0}}(q^{\bar{L}_0})\\ \non 
&=& \sum_{n\ge 0}\frac{q^{n^2}(-1)^n}{\eta(q)}\frac{q^{ks'^2}}{\eta (q)},
\label{aptr5}
\feqr 
\beqr 
q^{-1/12}Tr_{H^{Uv}_{0,0}}(q^{L_0}q^{\bar{L}_0})&=&q^{-1/24}Tr_{H^{0c}_{0,0}}(\frac{(1+g_c)}{2}q^{L_0})\cdot 
q^{-1/24}Tr_{H^{0\bar{c}}_{0,0}}(\frac{(1+g_a)}{2}q^{\bar{L}_0})\\ \non 
&=& \sum_{n\ge 0}\frac{q^{n^2}(-1)^n}{\eta(q)}\sum_{n\ge 0}\frac{q^{n^2}(-1)^n}{\eta(q)}.
\label{aptr6}
\feqr 
The proof of the rest of the lemmas proceeds similarly.

\addcontentsline{toc}{subsection}{Appendix D }
\section*{Appendix D }
\renewcommand{\theequation}{D.\arabic{equation}}
\setcounter{equation}{0}

To proceed with the proofs of the propositions we need first to derive the way the currents act on the lowest level states 
in our modules.  It is 
convenient to change the notation of our states in order to keep track of the charge.
So we rename the states as follows: 
\beqr
|s+s',k(s-s')+l>\rightarrow |s\sqrt{2k}+\frac{l}{\sqrt{2k}},s'\sqrt{2k}-\frac{l}{\sqrt{2k}}>.
\label{newnot}
\feqr
The free field representation of the currents gives  
\beqr 
&&\sum_{n\in Z}J_{-n}^{\pm }z^{n-k}|s\sqrt{2k}+\frac{l}{\sqrt{2k}},s'\sqrt{2k}-\frac{l}{\sqrt{2k}}>\\ \non 
&=& z^{\pm (2ks+l)}e^{\pm \sqrt{2k}\sum_{n>0}\frac{a_{-n}}{n}z^n}|(s\pm 1)\sqrt{2k}+\frac{l}{\sqrt{2k}},
s'\sqrt{2k}-\frac{l}{\sqrt{2k}}>, 
\\ \non 
&&\sum_{n\in Z}J_{-n}^{\pm }z^{n-k}|-s\sqrt{2k}-\frac{l}{\sqrt{2k}},s'\sqrt{2k}+\frac{l}{\sqrt{2k}}>\\ \non 
&=& z^{\mp (2ks+l)}e^{\pm \sqrt{2k}\sum_{n>0}\frac{a_{-n}}{n}z^n}|-(s\mp 1)\sqrt{2k}-\frac{l}{\sqrt{2k}},
s'\sqrt{2k}+\frac{l}{\sqrt{2k}}>.
\label{expncur1}
\feqr
Equating the lowest powers of z we get 
\beqr
\label{curact1} 
J^+_{-(2s+1)k-l}|s\sqrt{2k}+\frac{l}{\sqrt{2k}},s'\sqrt{2k}-\frac{l}{\sqrt{2k}}>&=&
|(s+1)\sqrt{2k}+\frac{l}{\sqrt{2k}},s'\sqrt{2k}-\frac{l}{\sqrt{2k}}>, \\ \non 
J^-_{-(2s+1)k-l}|-s\sqrt{2k}-\frac{l}{\sqrt{2k}},s'\sqrt{2k}+\frac{l}{\sqrt{2k}}>&=&
|-(s+1)\sqrt{2k}-\frac{l}{\sqrt{2k}},s'\sqrt{2k}+\frac{l}{\sqrt{2k}}>, \\ \non 
J^+_{(2s-1)k+l}|-s\sqrt{2k}-\frac{l}{\sqrt{2k}},s'\sqrt{2k}+\frac{l}{\sqrt{2k}}>&=&
|-(s-1)\sqrt{2k}-\frac{l}{\sqrt{2k}},s'\sqrt{2k}+\frac{l}{\sqrt{2k}}>, \\ \non 
J^-_{(2s-1)k+l}|s\sqrt{2k}+\frac{l}{\sqrt{2k}},s'\sqrt{2k}-\frac{l}{\sqrt{2k}}>&=&
|(s-1)\sqrt{2k}+\frac{l}{\sqrt{2k}},s'\sqrt{2k}-\frac{l}{\sqrt{2k}}>,
\feqr
while for the other powers of z we need to expand the exponentials in powers of z. This can be achieved conveniently 
using the Schur polynomials which are defined by the relation 
\beqr
e^{\sum_{n=1}^{\infty }t_nz^n}=\sum_{N=0}^{\infty }z^NS_N(t_1,t_2,\cdots ,t_N)
\label{schur1}
\feqr 
and turn out to be 
\beqr
\label{schur2} 
S_N(t_1,t_2,\cdots ,t_N)=\sum_{n_1,n_2,\cdots n_N}^{\sum_{i=1}^Nin_i=N}\frac{t_1^{n_1}\cdots t_N^{n_N}}{n_1!\cdots n_N!}. 
\feqr 
Using definition \rf{schur2} we get the following relations:
\beqr
\label{curact2} 
&& J^+_{-(2s+1)k-l}|-s\sqrt{2k}-\frac{l}{\sqrt{2k}},s'\sqrt{2k}+\frac{l}{\sqrt{2k}}> \\ \non 
&=& S_{4ks+2l}(\sqrt{2k}\frac{a_{-1}}{1},\sqrt{2k}\frac{a_{-2}}{2},\cdots ,\sqrt{2k}\frac{a_{-(4ks+2l)}}{(4ks+2l)!})
|-(s-1)\sqrt{2k}-\frac{l}{\sqrt{2k}},s'\sqrt{2k}+\frac{l}{\sqrt{2k}}>, \\ \non 
&&J^-_{-(2s+1)k-l}|s\sqrt{2k}+\frac{l}{\sqrt{2k}},s'\sqrt{2k}-\frac{l}{\sqrt{2k}}> \\ \non 
&=& S_{4ks+2l}(-\sqrt{2k}\frac{a_{-1}}{1},-\sqrt{2k}\frac{a_{-2}}{2},\cdots ,-\sqrt{2k}\frac{a_{-(4ks+2l)}}{(4ks+2l)!})
|(s-1)\sqrt{2k}+\frac{l}{\sqrt{2k}},s'\sqrt{2k}-\frac{l}{\sqrt{2k}}>.
\feqr 
The r.h.s. of the above equations vanishes unless $2ks+l>0$.  

{\it Proof of Proposition 1:} Starting from the vacuum state $|0,0>$ and acting on it with bilinears of 
the form $a_{-n_1}a_{-n_2}$ the space $H_{0,0}^{Uv}$ is generated.  To construct the remaining spaces 
we need to change the 
charge of the vacuum state. This can be accompished by acting with the currents. Consider 
\beqr 
(J^+_{-k}+J^-_{-k})|0,0>&=&|\sqrt{2k},0>+|-\sqrt{2k},0>, \\ \non 
a_{-n}(J^+_{-k}-J^-_{-k}) |0,0>&=&a_{-n}(|\sqrt{2k},0>-|-\sqrt{2k},0>). 
\label{modact1}
\feqr 
The bilinears $a_{-n_1}a_{-n_2}$ and their conjugates acting on these states are sufficient to generate 
the module $H^{Uv}_{1,k}$.  Proceeding inductively, suppose we have generated the spaces $H^{Uv}_{s-1,k(s-1)}$  
$H^{Uv}_{s,ks}$.  
Acting on the state $|s\sqrt{2k},0>+|-s\sqrt{2k},0>$ with $J^+_{-(2s+1)k}+J^-_{-(2s+1)k}$ and 
$a_{-n}(J^+_{-(2s+1)k}-J^-_{-(2s+1)k})$ we get 
\beqr 
&&(J^+_{-(2s+1)k}+J^-_{-(2s+1)k})(|s\sqrt{2k},0>+|-s\sqrt{2k},0>)\\ \non 
&=&|(s+1)\sqrt{2k},0>+|-(s+1)\sqrt{2k},0> \\ \non 
&+&S_{4ks}(\sqrt{2k}\frac{a_{-1}}{1},\sqrt{2k}\frac{a_{-2}}{2},\cdots ,\sqrt{2k}\frac{a_{-4ks}}{(4ks)!})|-(s-1)\sqrt{2k},0>
\\ \non 
&+& S_{4ks}(-\sqrt{2k}\frac{a_{-1}}{1},-\sqrt{2k}\frac{a_{-2}}{2},\cdots ,-\sqrt{2k}\frac{a_{-4ks}}{(4ks)!})
|(s-1)\sqrt{2k},0>\\ \non 
&=& |(s+1)\sqrt{2k},0>+|-(s+1)\sqrt{2k},0>\\ \non 
&+& S^e_{4ks}(\sqrt{2k}\frac{a_{-1}}{1},\sqrt{2k}\frac{a_{-2}}{2},\cdots ,\sqrt{2k}\frac{a_{-4ks}}{(4ks)!})
(|(s-1)\sqrt{2k},0>+|-(s-1)\sqrt{2k},0>)\\ \non 
&-&S^o_{4ks}(\sqrt{2k}\frac{a_{-1}}{1},\sqrt{2k}\frac{a_{-2}}{2},\cdots ,\sqrt{2k}\frac{a_{-4k)}}{(4ks)!})
(|(s-1)\sqrt{2k},0>-|-(s-1)\sqrt{2k},0>)\\ \non 
&=& (|(s+1)\sqrt{2k},0>+|-(s+1)\sqrt{2k},0>)modH^{Uv}_{s-1,k(s-1)},
\label{modind1}
\feqr
\beqr 
&&a_{-n}(J^+_{-(2s+1)k}-J^-_{-(2s+1)k})(|s\sqrt{2k},0>+|-s\sqrt{2k},0>)\\ \non 
&=&a_{-n}(|(s+1)\sqrt{2k},0>-|-(s+1)\sqrt{2k},0>) \\ \non 
&+&a_{-n}S_{4ks}(\sqrt{2k}\frac{a_{-1}}{1},\sqrt{2k}\frac{a_{-2}}{2},\cdots , \sqrt{2k}\frac{a_{-4ks}}{(4ks)!})
|-(s-1)\sqrt{2k},0> \\ \non 
&-& a_{-n}S_{4ks}(-\sqrt{2k}\frac{a_{-1}}{1},-\sqrt{2k}\frac{a_{-2}}{2},\cdots ,-\sqrt{2k}\frac{a_{-4ks}}{(4ks)!})
|(s-1)\sqrt{2k},0>\\ \non 
&=& a_{-n}(|(s+1)\sqrt{2k},0>-|-(s+1)\sqrt{2k},0>)\\ \non 
&+&a_{-n} S^o_{4ks}(\sqrt{2k}\frac{a_{-1}}{1},\sqrt{2k}\frac{a_{-2}}{2},\cdots ,\sqrt{2k}\frac{a_{-4ks}}{(4ks)!})
(|(s-1)\sqrt{2k},0>+|-(s-1)\sqrt{2k},0>)\\ \non 
&-&a_{-n}S^e_{4ks}(\sqrt{2k}\frac{a_{-1}}{1},\sqrt{2k}\frac{a_{-2}}{2},\cdots ,\sqrt{2k}\frac{a_{-4ks}}{(4ks)!})
(|(s-1)\sqrt{2k},0>-|-(s-1)\sqrt{2k},0>)\\ \non 
&=& a_{-n}(|(s+1)\sqrt{2k},0>-|-(s+1)\sqrt{2k},0>)modH^{Uv}_{s-1,k(s-1)}
\label{modind2}
\feqr
where $S^e_{4ks}$ and $S^o_{4ks}$ are the parts of the Schur polynomials that contain products of even or odd  
numbers of oscillator modes respectively.  In this way we have generated the lowest level states of the module 
$H^{Uv}_{s+1,k(s+1)}$. Acting on 
these states with the bilinears of modes we get the whole  $H^{Uv}_{s+1,k(s+1)}$. Thus the induction step is complete and 
we have generated all the modules $H^{Uv}_{s,ks}$ starting from the vacuum state . 
\par Similarly we get $H^{Uv}_{s',-ks'}$.  Before embarking on induction to get the spaces $H^{Uv}_{s+s',k(s-s')}$ we 
need to construct explicitly the space $H^{Uv}_{1+1,k(1-1)}=H^{Uv}_{2,0}$.  This can be constructed from the 
bilinears of modes and the relations:
\beqr 
(J^+_{-k}+J^-_{-k})(\bar{J}^+_{-k}+\bar{J}^-_{-k})|0,0>&=& |\sqrt{2k},\sqrt{2k}> +|\sqrt{2k},-\sqrt{2k}> \\ \non 
&+& |-\sqrt{2k},\sqrt{2k}> +|-\sqrt{2k},-\sqrt{2k}> \\ \non 
a_{-n_1}(J^+_{-k}-J^-_{-k})\bar{a}_{-n_2}(\bar{J}^+_{-k}-\bar{J}^-_{-k})|0,0>&=&
a_{-n_1}\bar{a}_{-n_2}( |\sqrt{2k},\sqrt{2k}> -|\sqrt{2k},-\sqrt{2k}> \\ \non
&-& |-\sqrt{2k},\sqrt{2k}>+|-\sqrt{2k},-\sqrt{2k}>)
\\ \non 
a_{-n}(J^+_{-k}-J^-_{-k})(\bar{J}^+_{-k}+\bar{J}^-_{-k})|0,0>&=& 
a_{-n}( |\sqrt{2k},\sqrt{2k}> +|\sqrt{2k},-\sqrt{2k}> \\ \non
&-& |-\sqrt{2k},\sqrt{2k}>-|-\sqrt{2k},-\sqrt{2k}>) \\ \non 
(J^+_{-k}+J^-_{-k})\bar{a}_{-n}(\bar{J}^+_{-k}-\bar{J}^-_{-k})|0,0>&=&
\bar{a}_{-n}( |\sqrt{2k},\sqrt{2k}> -|\sqrt{2k},-\sqrt{2k}> \\ \non
&+& |-\sqrt{2k},\sqrt{2k}>-|-\sqrt{2k},-\sqrt{2k}>).
\label{modact2}
\feqr 
\par The inductive step is as follows:  
Assuming that we have constructed all the spaces $H^{Uv}_{s+s',k(s-s')}$ for $s,s'\le p$ where $p\ge 1$ we 
will prove that we can construct all the spaces $H^{Uv}_{s+s',k(s-s')}$ with $s,s'\le p+1$.  To show this 
we need to construct the spaces $H^{Uv}_{(p+1)+s',k((p+1)-s')}$, $H^{Uv}_{s+(p+1),k(s-(p+1))}$ and 
$H^{Uv}_{2p+2,0}$.  To build up the first spaces we need the relations 
\beqr 
&&(J^+_{-(2p+1)k}+J^-_{-(2p+1)k}) (|p\sqrt{2k},s'\sqrt{2k}>+ |-p\sqrt{2k},s'\sqrt{2k}> \\ \non
&+&|p\sqrt{2k},-s'\sqrt{2k}> +|-p\sqrt{2k},-s'\sqrt{2k}> )\\ \non 
&=& (|(p+1)\sqrt{2k},s'\sqrt{2k}>+ |-(p+1)\sqrt{2k},s'\sqrt{2k}> \\ \non
&+& |(p+1)\sqrt{2k},-s'\sqrt{2k}> +|-(p+1)\sqrt{2k},-s'\sqrt{2k}> )modH^{Uv}_{p-1+s',k((p-1)-s')},\\ \non 
\\ \non
&&a_{-n}(J^+_{-(2p+1)k}-J^-_{-(2p+1)k}) (|p\sqrt{2k},s'\sqrt{2k}>+ |-p\sqrt{2k},s'\sqrt{2k}> \\ \non
&+& |p\sqrt{2k},-s'\sqrt{2k}> +|-p\sqrt{2k},-s'\sqrt{2k}> )\\ \non 
&=&a_{-n}(|(p+1)\sqrt{2k},s'\sqrt{2k}>- |-(p+1)\sqrt{2k},s'\sqrt{2k}> \\ \non
&+& |(p+1)\sqrt{2k},-s'\sqrt{2k}> -|-(p+1)\sqrt{2k},-s'\sqrt{2k}> )modH^{Uv}_{p-1+s',k((p-1)-s')},\\ \non 
\\ \non
&&(J^+_{-(2p+1)k}+J^-_{-(2p+1)k}) \bar{a}_{-n} (|p\sqrt{2k},s'\sqrt{2k}>+ |-p\sqrt{2k},s'\sqrt{2k}> \\ \non
&-& |p\sqrt{2k},-s'\sqrt{2k}> -|-p\sqrt{2k},-s'\sqrt{2k}> )\\ \non 
&=& \bar{a}_{-n}(|(p+1)\sqrt{2k},s'\sqrt{2k}>+ |-(p+1)\sqrt{2k},s'\sqrt{2k}> \\ \non
&-& |(p+1)\sqrt{2k},-s'\sqrt{2k}> -|-(p+1)\sqrt{2k},-s'\sqrt{2k}> )modH^{Uv}_{p-1+s',k((p-1)-s')},\\ \non 
\\ \non
&&a_{-n_1}(J^+_{-(2p+1)k}-J^-_{-(2p+1)k}) \bar{a}_{-n_2} (|p\sqrt{2k},s'\sqrt{2k}>+ |-p\sqrt{2k},s'\sqrt{2k}> \\ \non
&-& |p\sqrt{2k},-s'\sqrt{2k}> -|-p\sqrt{2k},-s'\sqrt{2k}> )\\ \non 
&=& a_{-n_1}\bar{a}_{-n_2}(|(p+1)\sqrt{2k},s'\sqrt{2k}>-|-(p+1)\sqrt{2k},s'\sqrt{2k}> \\ \non
&-& |(p+1)\sqrt{2k},-s'\sqrt{2k}> +|-(p+1)\sqrt{2k},-s'\sqrt{2k}> )modH^{Uv}_{p-1+s',k((p-1)-s')}, 
\label{modind3}
\feqr
and the usual bilinears of modes.  The construction of the second spaces proceeds similarly.
 The third space can be constructed from   
$H^{Uv}_{(p-1)+(p+1),k((p-1)-(p+1))}$ and $H^{Uv}_{p+(p+1),k(p-(p+1))}$ using the above relations.  

\par So we have shown that the space $H_0$ can be constructed starting from the vacuum state 
and acting on it by the operators in S. Extend now the definition of the operators $g_c$, $g_a$ defined in Appendix C, by 
making them send $s$ to $-s$ and $s'$ to $-s'$ respectively:
\beqr 
g_c(|s\sqrt{2k},s'\sqrt{2k}>=|-s\sqrt{2k},s'\sqrt{2k}>, \\ \non 
g_a(|s\sqrt{2k},s'\sqrt{2k}>=|s\sqrt{2k},-s'\sqrt{2k}>.
\label{defgc}
\feqr
It is an easy excercise to check that these operators commute with the action of the elements of S.  
Since the elements of S change the conformal dimension of the 
states by integers, the space they generate starting from the vacuum is a subspace of $H_0\oplus H_1$.  
But the space $H_0$ is the eigenspace of $g_c$ with eigenvalue $+1$ (also the 
eigenspace of $g_a$ with eigenvalue $+1$) while  
$H_1$ is the eigenspace of $g_c$ (resp. of $g_a$) with eigenvalue $-1$ (resp. $-1$). Since the generators 
of S commute with $g_c$ (resp. $g_a$) our space is a subspace of $H_0$.  But since $H_0$ is generated from the vacuum, 
the action of S on the vacuum generates precisely $H_0$.  

{\it Proof of Proposition 2:} The proof of proposition 2 proceeds along the same lines as proposition 1 with the only 
difference that equations \rf{curact1} are replaced by
\beqr 
J^+_{-(2s+1)k+1} a_{-1} \bar{a}_{-1} |s\sqrt{2k},s'\sqrt{2k}>&=&
-\sqrt{2k}\bar{a}_{-1}|(s+1)\sqrt{2k},s'\sqrt{2k}>, \\ \non 
J^-_{-(2s+1)k+1} a_{-1} \bar{a}_{-1} |-s\sqrt{2k},s'\sqrt{2k}>&=&
\sqrt{2k} \bar{a}_{-1}|-(s+1)\sqrt{2k},s'\sqrt{2k}>
\label{curact5}
\feqr
and equations \rf{curact2} are replaced by 
\beqr 
&&J^+_{-(2s+1)k+1}a_{-1} \bar{a}_{-1} |-s\sqrt{2k},s'\sqrt{2k}> \\ \non 
&=&\Bigl[ -\sqrt{2k} S_{4ks}(\sqrt{2k}\frac{a_{-1}}{1},\sqrt{2k}\frac{a_{-2}}{2},\cdots ,\sqrt{2k}\frac{a_{-4ks}}{(4ks)!})
\\ \non 
&+& S_{4ks-1}(\sqrt{2k}\frac{a_{-1}}{1},\sqrt{2k}\frac{a_{-2}}{2},\cdots ,\sqrt{2k}\frac{a_{-4ks+1}}{(4ks-1)!})a_{-1}
\Bigr] \, \bar{a}_{-1}|-(s-1)\sqrt{2k},s'\sqrt{2k}>,\\ \non 
\\ \non 
&&J^-_{-(2s+1)k+1}a_{-1} \bar{a}_{-1}|s\sqrt{2k},s'\sqrt{2k}> \\ \non 
&=& \Bigl[ \sqrt{2k} S_{4ks}(-\sqrt{2k}\frac{a_{-1}}{1},-\sqrt{2k}\frac{a_{-2}}{2},\cdots ,-
\sqrt{2k}\frac{a_{-4ks}}{(4ks)!}) \\ \non 
&+& S_{4ks-1}(-\sqrt{2k}\frac{a_{-1}}{1},-\sqrt{2k}\frac{a_{-2}}{2},\cdots ,-
\sqrt{2k}\frac{a_{-4ks+1}}{(4ks-1)!})a_{-1} \Bigr] \, \bar{a}_{-1}|(s-1)\sqrt{2k},s'\sqrt{2k}>
\label{curact6}
\feqr 

{\it Proof of Proposition 3:} In our new notation the state $|0,l>+|0,-l>$ becomes 
$ |l/\sqrt{2k},-l/\sqrt{2k}>+|-l/\sqrt{2k},l/\sqrt{2k}> $. It is also 
convenient to change the notation of the modules as follows:
\beqr 
H^{Ul}_{s,s'}\equiv H^U_{s+s',k(s-s')+l}.
\label{notat}
\feqr
Now we have
\scriptsize  
\beqr 
&&a_{-n}(J^+_{k-l}-J^-_{k-l})(J^+_{-k+l}+J^-_{-k+l})|\frac{l}{\sqrt{2k}},-\frac{l}{\sqrt{2k}}>
+|-\frac{l}{\sqrt{2k}},\frac{l}{\sqrt{2k}}> 
=a_{-n} (|\frac{l}{\sqrt{2k}},-\frac{l}{\sqrt{2k}}>-|-\frac{l}{\sqrt{2k}},\frac{l}{\sqrt{2k}}>), 
\\ \non 
\\ \non 
\\ \non
&&-\bar{a}_{-n}(\bar{J}^+_{k-l}-\bar{J}^-_{k-l})(\bar{J}^+_{-k+l}+\bar{J}^-_{-k+l})
(|\frac{l}{\sqrt{2k}},-\frac{l}{\sqrt{2k}}>+|-\frac{l}{\sqrt{2k}},\frac{l}{\sqrt{2k}}>)  
=\bar{a}_{-n}(|\frac{l}{\sqrt{2k}},-\frac{l}{\sqrt{2k}}>-|-\frac{l}{\sqrt{2k}},\frac{l}{\sqrt{2k}}>), 
\\ \non 
\\ \non 
\\ \non
&&-\bar{a}_{-n_2}(\bar{J}^+_{k-l}-\bar{J}^-_{k-l})a_{-n_1}(J^+_{k-l}-J^-_{k-l}) 
(\bar{J}^+_{-k+l}+\bar{J}^-_{-k+l})(J^+_{-k+l}+J^-_{-k+l})  
(|\frac{l}{\sqrt{2k}},-\frac{l}{\sqrt{2k}}>+|-\frac{l}{\sqrt{2k}},\frac{l}{\sqrt{2k}}>) \\ \non 
&=&\bar{a}_{-n_2}a_{-n_1}(|\frac{l}{\sqrt{2k}},-\frac{l}{\sqrt{2k}}>+|-\frac{l}{\sqrt{2k}},\frac{l}{\sqrt{2k}}>).
\label{lowst1}
\feqr
\normalsize 
These states together with the bilinears of modes are sufficient to construct $H^{Ul}_{0,0}$.  Also we have  
\beqr 
&&(J^+_{-k+l}+J^-_{-k+l})(|\frac{l}{\sqrt{2k}},-\frac{l}{\sqrt{2k}}>+|-\frac{l}{\sqrt{2k}},\frac{l}{\sqrt{2k}}>)\\ \non 
&=&(|-\sqrt{2k}+\frac{l}{\sqrt{2k}},-\frac{l}{\sqrt{2k}}>+|\sqrt{2k}-\frac{l}{\sqrt{2k}},\frac{l}{\sqrt{2k}}>), 
\\ \non 
\\ \non 
&&-a_{-n}(J^+_{-k+l}-J^-_{-k+l})(|\frac{l}{\sqrt{2k}},-\frac{l}{\sqrt{2k}}>+|-\frac{l}{\sqrt{2k}},\frac{l}{\sqrt{2k}}>)
\\ \non 
&=&a_{-n}(|-\sqrt{2k}+\frac{l}{\sqrt{2k}},-\frac{l}{\sqrt{2k}}>-|\sqrt{2k}-\frac{l}{\sqrt{2k}},\frac{l}{\sqrt{2k}}>),
\\ \non 
\\ \non 
&&(J^+_{-k+l}+J^-_{-k+l})\bar{a}_{-n}(|\frac{l}{\sqrt{2k}},-\frac{l}{\sqrt{2k}}>
-|-\frac{l}{\sqrt{2k}},\frac{l}{\sqrt{2k}}>)\\ \non 
&=&\bar{a}_{-n}(|-\sqrt{2k}+\frac{l}{\sqrt{2k}},-\frac{l}{\sqrt{2k}}>
-|\sqrt{2k}-\frac{l}{\sqrt{2k}},\frac{l}{\sqrt{2k}}>), 
\\ \non 
\\ \non 
&&-a_{-n_1}(J^+_{-k+l}-J^-_{-k+l})\bar{a}_{-n_2}(|\frac{l}{\sqrt{2k}},-\frac{l}{\sqrt{2k}}>
-|-\frac{l}{\sqrt{2k}},\frac{l}{\sqrt{2k}}>)\\ \non 
&=&a_{-n_1}\bar{a}_{-n_2}(|-\sqrt{2k}+\frac{l}{\sqrt{2k}},-\frac{l}{\sqrt{2k}}>
+|\sqrt{2k}-\frac{l}{\sqrt{2k}},\frac{l}{\sqrt{2k}}>)
\label{lowst2}
\feqr
and, together with the bilinears of the modes, we are able to construct $H^{Ul}_{-1,0}$.  Similarly 
we can construct the spaces $H^{Ul}_{-1,-1}$ and $H^{Ul}_{0,-1}$.  

\par Now we have to proceed inductively on s.  Recall that the spaces $H^{Ul}_{-1,0}$, 
$H^{Ul}_{-1,-1}$, $H^{Ul}_{0,-1}$ and $H^{Ul}_{0,0}$ have been constructed.  
Assuming that the spaces $H^{Ul}_{s,s'}$ for $-1\le s,s'\le p$ have been constructed we will construct the spaces 
$H^{Ul}_{p+1,s'}$, $H^{Ul}_{s,p+1}$ and $H^{Ul}_{p+1,p+1}$. For the spaces $H^{Ul}_{p+1,s'}$ we need the following
relations:
\scriptsize
\beqr 
&&(J^+_{-(2p+1)k-l}+J^-_{-(2p+1)k-l}) (|p\sqrt{2k}+\frac{l}{\sqrt{2k}},s'\sqrt{2k}-\frac{l}{\sqrt{2k}}>+
|-p\sqrt{2k}-\frac{l}{\sqrt{2k}},-s'\sqrt{2k}+\frac{l}{\sqrt{2k}}>) \\ \non 
&=&(|(p+1)\sqrt{2k}+\frac{l}{\sqrt{2k}},s'\sqrt{2k}-\frac{l}{\sqrt{2k}}>
+|-(p+1)\sqrt{2k}-\frac{l}{\sqrt{2k}},-s'\sqrt{2k}+\frac{l}{\sqrt{2k}}>)modH^{Ul}_{p-1,s'}, \\ \non 
\\ \non 
&&a_{-n}(J^+_{-(2p+1)k-l}-J^-_{-(2p+1)k-l}) (|p\sqrt{2k}+\frac{l}{\sqrt{2k}},s'\sqrt{2k}-\frac{l}{\sqrt{2k}}>+
|-p\sqrt{2k}-\frac{l}{\sqrt{2k}},-s'\sqrt{2k}+\frac{l}{\sqrt{2k}}>) \\ \non 
&=& a_{-n}(|(p+1)\sqrt{2k}+\frac{l}{\sqrt{2k}},s'\sqrt{2k}-\frac{l}{\sqrt{2k}}>
-|-(p+1)\sqrt{2k}-\frac{l}{\sqrt{2k}},-s'\sqrt{2k}+\frac{l}{\sqrt{2k}}>)modH^{Ul}_{p-1,s'},\\ \non 
\\ \non 
&&(J^+_{-(2p+1)k-l}+J^-_{-(2p+1)k-l}) \bar{a}_{-n}(|p\sqrt{2k}+\frac{l}{\sqrt{2k}},s'\sqrt{2k}-\frac{l}{\sqrt{2k}}>-
|-p\sqrt{2k}-\frac{l}{\sqrt{2k}},-s'\sqrt{2k}+\frac{l}{\sqrt{2k}}>) \\ \non 
&=&\bar{a}_{-n}(|(p+1)\sqrt{2k}+\frac{l}{\sqrt{2k}},s'\sqrt{2k}-\frac{l}{\sqrt{2k}}>
-|-(p+1)\sqrt{2k}-\frac{l}{\sqrt{2k}},-s'\sqrt{2k}+\frac{l}{\sqrt{2k}}>)modH^{Ul}_{p-1,s'}, \\ \non 
\\ \non 
&&a_{-n_1}(J^+_{-(2p+1)k-l}-J^-_{-(2p+1)k-l})\bar{a}_{-n_2} (|p\sqrt{2k}+\frac{l}{\sqrt{2k}},s'\sqrt{2k}-\frac{l}{\sqrt{2k}}>-
|-p\sqrt{2k}-\frac{l}{\sqrt{2k}},-s'\sqrt{2k}+\frac{l}{\sqrt{2k}}>) \\ \non 
&=& a_{-n_1}\bar{a}_{-n_2}(|(p+1)\sqrt{2k}+\frac{l}{\sqrt{2k}},s'\sqrt{2k}-\frac{l}{\sqrt{2k}}>+
|-(p+1)\sqrt{2k}-\frac{l}{\sqrt{2k}},-s'\sqrt{2k}+\frac{l}{\sqrt{2k}}>)modH^{Ul}_{p-1,s'}.
\label{modind10}
\feqr
\normalsize 
Together with the bilinears of modes we can construct the spaces $H^{Ul}_{p+1,s'}$. 
Similarly we can construct the spaces $H^{Ul}_{s,p+1}$,  while the space $H^{Ul}_{p+1,p+1}$ can be constructed from 
the spaces $H^{Ul}_{p-1,p+1}$ and $H^{Ul}_{p,p+1}$ using the above relations.  So finally we have constructed all the 
spaces $H^{Ul}_{s,s'}$ with $s,s'\ge -1$.  

\par To proceed further we need induction on the negative values of $s,s'$.  Assuming again that all the spaces 
$H^{Ul}_{s,s'}$ for $s,s'\ge -p$ have been constructed we will show that we can construct the spaces 
$H^{Ul}_{-p-1,s'}$, $H^{Ul}_{s,-p-1}$ and $H^{Ul}_{-p-1,-p-1}$.  For the spaces $H^{Ul}_{-p-1,s'}$ we need the relations 
\scriptsize
\beqr 
&&(J^+_{-(2p+1)k+l}+J^-_{-(2p+1)k+l}) (|-p\sqrt{2k}+\frac{l}{\sqrt{2k}},s'\sqrt{2k}-\frac{l}{\sqrt{2k}}>+
|p\sqrt{2k}-\frac{l}{\sqrt{2k}},-s'\sqrt{2k}+\frac{l}{\sqrt{2k}}>) \\ \non 
&=&(|-(p+1)\sqrt{2k}+\frac{l}{\sqrt{2k}},s'\sqrt{2k}-\frac{l}{\sqrt{2k}}>+
|(p+1)\sqrt{2k}-\frac{l}{\sqrt{2k}},-s'\sqrt{2k}+\frac{l}{\sqrt{2k}}>)modH^{Ul}_{-p+1,s'}, \\ \non 
\\ \non 
&&-a_{-n}(J^+_{-(2p+1)k+l}-J^-_{-(2p+1)k+l}) (|-p\sqrt{2k}+\frac{l}{\sqrt{2k}},s'\sqrt{2k}-\frac{l}{\sqrt{2k}}>+
|p\sqrt{2k}-\frac{l}{\sqrt{2k}},-s'\sqrt{2k}+\frac{l}{\sqrt{2k}}>) \\ \non 
&=& a_{-n}(|-(p+1)\sqrt{2k}+\frac{l}{\sqrt{2k}},s'\sqrt{2k}-\frac{l}{\sqrt{2k}}>
-|(p+1)\sqrt{2k}-\frac{l}{\sqrt{2k}},-s'\sqrt{2k}+\frac{l}{\sqrt{2k}}>)modH^{Ul}_{-p+1,s'},\\ \non 
\\ \non 
&&(J^+_{-(2p+1)k+l}+J^-_{-(2p+1)k+l}) \bar{a}_{-n}(|-p\sqrt{2k}+\frac{l}{\sqrt{2k}},s'\sqrt{2k}-\frac{l}{\sqrt{2k}}>-
|p\sqrt{2k}-\frac{l}{\sqrt{2k}},-s'\sqrt{2k}+\frac{l}{\sqrt{2k}}>) \\ \non 
&=&\bar{a}_{-n}(|-(p+1)\sqrt{2k}+\frac{l}{\sqrt{2k}},s'\sqrt{2k}-\frac{l}{\sqrt{2k}}>
-|(p+1)\sqrt{2k}-\frac{l}{\sqrt{2k}},-s'\sqrt{2k}+\frac{l}{\sqrt{2k}}>)modH^{Ul}_{-p+1,s'}, \\ \non 
\\ \non 
&&-a_{-n_1}(J^+_{-(2p+1)k+l}-J^-_{-(2p+1)k+l})\bar{a}_{-n_2} (|-p\sqrt{2k}
+\frac{l}{\sqrt{2k}},s'\sqrt{2k}-\frac{l}{\sqrt{2k}}>-
|p\sqrt{2k}-\frac{l}{\sqrt{2k}},-s'\sqrt{2k}+\frac{l}{\sqrt{2k}}>) \\ \non 
&=& a_{-n_1}\bar{a}_{-n_2}(|-(p+1)\sqrt{2k}+\frac{l}{\sqrt{2k}},s'\sqrt{2k}-\frac{l}{\sqrt{2k}}>+
|(p+1)\sqrt{2k}-\frac{l}{\sqrt{2k}},-s'\sqrt{2k}+\frac{l}{\sqrt{2k}}>)modH^{Ul}_{-p+1,s'}
\label{modind11}
\feqr
\normalsize 
and the bilinears of the modes.  The second and third spaces are constructed similarly.  
So we have shown that starting from the state $ |l/\sqrt{2k},-l/\sqrt{2k}>+|-l/\sqrt{2k},l/\sqrt{2k}> $ 
we can construct $H_{l^2/4k}=\oplus_{s,s'\in Z}H^{Ul}_{s,s'}$.  

{\it Proof of Proposition 4:} Here we will denote $H^{Uvk}_{s,s'}\equiv H^{Uv}_{s+s'+1,k(s-s')}$ and 
$H^{Ubk}_{s,s'}\equiv H^{Ub}_{s+s'+1,k(s-s')}$.  In the new notation the relevant states become:
\beqr 
&&|s+s'+1,k(s-s')>=|s+(s'+1),k(s-(s'+1))+k> \\ \non
&\rightarrow & |s\sqrt{2k}+ \frac{k}{\sqrt{2k}},(s'+1)\sqrt{2k}-\frac{k}{\sqrt{2k}}>=
|(s+1/2)\sqrt{2k},(s'+1/2)\sqrt{2k}>, \\ \non 
\\ \non
&&|-(s+s'+1),-k(s-s')>=|-(s+(s'+1)),-k(s-(s'+1))-k> \\ \non
&\rightarrow & |-s\sqrt{2k}-\frac{k}{\sqrt{2k}},-(s'+1)\sqrt{2k}+\frac{k}{\sqrt{2k}}>=
|-(s+1/2)\sqrt{2k},-(s'+1/2)\sqrt{2k}>, \\ \non 
\\ \non
&&|s-s',k(s+s'+1)>=|s-s',k(s+s')+k> \\ \non
&\rightarrow & |s\sqrt{2k}+\frac{k}{\sqrt{2k}},-s'\sqrt{2k}-\frac{k}{\sqrt{2k}}>=
|(s+1/2)\sqrt{2k},-(s'+1/2)\sqrt{2k}>, \\ \non 
\\ \non
&&|-(s-s'),-k(s+s'+1)>=|-(s-s'),-k(s+s')-k> \\ \non
&\rightarrow & |-s\sqrt{2k}-\frac{k}{\sqrt{2k}},s'\sqrt{2k}+\frac{k}{\sqrt{2k}}>=
|-(s+1/2)\sqrt{2k},(s'+1/2)\sqrt{2k}>. 
\label{stat5}
\feqr
We have to prove that starting from the states 
\beqr 
|\frac{1}{2}\sqrt{2k},\frac{1}{2}\sqrt{2k}> + |-\frac{1}{2}\sqrt{2k},-\frac{1}{2}\sqrt{2k}>
+|\frac{1}{2}\sqrt{2k},-\frac{1}{2}\sqrt{2k}>+|-\frac{1}{2}\sqrt{2k},\frac{1}{2}\sqrt{2k}>, \\ \non 
|\frac{1}{2}\sqrt{2k},\frac{1}{2}\sqrt{2k}> + |-\frac{1}{2}\sqrt{2k},-\frac{1}{2}\sqrt{2k}>
-|\frac{1}{2}\sqrt{2k},-\frac{1}{2}\sqrt{2k}>-|-\frac{1}{2}\sqrt{2k},\frac{1}{2}\sqrt{2k}> 
\label{stat6}
\feqr 
we can generate the modules $H^v_{k/4}$ and $H^b_{k/4}$ respectively.  We show only the construction of $H^v_{k/4}$ 
since the construction of $H^b_{k/4}$ proceeds similarly.  To generate $H^{Uvk}_{0,0}$ we need the 
following relations:  
\scriptsize
\beqr 
&& a_{-n}(J^+_0-J^-_0) (|\frac{1}{2}\sqrt{2k},\frac{1}{2}\sqrt{2k}> + |-\frac{1}{2}\sqrt{2k},-\frac{1}{2}\sqrt{2k}>
+|\frac{1}{2}\sqrt{2k},-\frac{1}{2}\sqrt{2k}>+|-\frac{1}{2}\sqrt{2k},\frac{1}{2}\sqrt{2k}>) \\ \non 
&=&a_{-n}(|\frac{1}{2}\sqrt{2k},\frac{1}{2}\sqrt{2k}> + |\frac{1}{2}\sqrt{2k},-\frac{1}{2}\sqrt{2k}>
-|-\frac{1}{2}\sqrt{2k},\frac{1}{2}\sqrt{2k}>-|-\frac{1}{2}\sqrt{2k},-\frac{1}{2}\sqrt{2k}>), \\ \non 
\\ \non
&& \bar{a}_{-n}(\bar{J}^+_0-\bar{J}^-_0)
(|\frac{1}{2}\sqrt{2k},\frac{1}{2}\sqrt{2k}> + |-\frac{1}{2}\sqrt{2k},-\frac{1}{2}\sqrt{2k}>
+|\frac{1}{2}\sqrt{2k},-\frac{1}{2}\sqrt{2k}>+|-\frac{1}{2}\sqrt{2k},\frac{1}{2}\sqrt{2k}>) \\ \non 
&=&\bar{a}_{-n}(|\frac{1}{2}\sqrt{2k},\frac{1}{2}\sqrt{2k}> + |-\frac{1}{2}\sqrt{2k},\frac{1}{2}\sqrt{2k}>
-|\frac{1}{2}\sqrt{2k},-\frac{1}{2}\sqrt{2k}>-|-\frac{1}{2}\sqrt{2k},-\frac{1}{2}\sqrt{2k}>), \\ \non 
\\ \non
&& a_{-n_1}(J^+_0-J^-_0)\bar{a}_{-n_2}(\bar{J}^+_0-\bar{J}^-_0)
(|\frac{1}{2}\sqrt{2k},\frac{1}{2}\sqrt{2k}> + |-\frac{1}{2}\sqrt{2k},-\frac{1}{2}\sqrt{2k}>
+|\frac{1}{2}\sqrt{2k},-\frac{1}{2}\sqrt{2k}>+|-\frac{1}{2}\sqrt{2k},\frac{1}{2}\sqrt{2k}>) \\ \non 
&=&a_{-n_1}\bar{a}_{-n_2}(|\frac{1}{2}\sqrt{2k},\frac{1}{2}\sqrt{2k}> -|\frac{1}{2}\sqrt{2k},-\frac{1}{2}\sqrt{2k}>
-|-\frac{1}{2}\sqrt{2k},\frac{1}{2}\sqrt{2k}>+|-\frac{1}{2}\sqrt{2k},-\frac{1}{2}\sqrt{2k}>). 
\label{modind12}
\feqr
\normalsize 
To generate $H^{Uvk}_{1,0}$ we need the following relations:
\scriptsize
\beqr 
&& (J^+_{-2k}+J^-_{-2k}) (|\frac{1}{2}\sqrt{2k},\frac{1}{2}\sqrt{2k}> + |-\frac{1}{2}\sqrt{2k},\frac{1}{2}\sqrt{2k}>
+|\frac{1}{2}\sqrt{2k},-\frac{1}{2}\sqrt{2k}>+|-\frac{1}{2}\sqrt{2k},-\frac{1}{2}\sqrt{2k}>) \\ \non 
&=& (|(1+\frac{1}{2})\sqrt{2k},\frac{1}{2}\sqrt{2k}> + |-(1+\frac{1}{2})\sqrt{2k},\frac{1}{2}\sqrt{2k}>
+|(1+\frac{1}{2})\sqrt{2k},-\frac{1}{2}\sqrt{2k}>+|-(1+\frac{1}{2})\sqrt{2k},-\frac{1}{2}\sqrt{2k}>)
modH^{Uvk}_{0,0}, \\ \non  
\\ \non 
&& a_{-n}(J^+_{-2k}-J^-_{-2k})(|\frac{1}{2}\sqrt{2k},\frac{1}{2}\sqrt{2k}> + |-\frac{1}{2}\sqrt{2k},\frac{1}{2}\sqrt{2k}>
+|\frac{1}{2}\sqrt{2k},-\frac{1}{2}\sqrt{2k}>+|-\frac{1}{2}\sqrt{2k},-\frac{1}{2}\sqrt{2k}>) \\ \non  
&=&a_{-n} (|(1+\frac{1}{2})\sqrt{2k},\frac{1}{2}\sqrt{2k}> - |-(1+\frac{1}{2})\sqrt{2k},\frac{1}{2}\sqrt{2k}>
+|(1+\frac{1}{2})\sqrt{2k},-\frac{1}{2}\sqrt{2k}>-|-(1+\frac{1}{2})\sqrt{2k},-\frac{1}{2}\sqrt{2k}>)
modH^{Uvk}_{0,0}, \\ \non  
\\ \non 
&& (J^+_{-2k}+J^-_{-2k}) \bar{a}_{-n} (|\frac{1}{2}\sqrt{2k},\frac{1}{2}\sqrt{2k}> + |-\frac{1}{2}\sqrt{2k},\frac{1}{2}\sqrt{2k}>
-|\frac{1}{2}\sqrt{2k},-\frac{1}{2}\sqrt{2k}>-|-\frac{1}{2}\sqrt{2k},-\frac{1}{2}\sqrt{2k}>) \\ \non 
&=& \bar{a}_{-n}(|(1+\frac{1}{2})\sqrt{2k},\frac{1}{2}\sqrt{2k}> + |-(1+\frac{1}{2})\sqrt{2k},\frac{1}{2}\sqrt{2k}>
-|(1+\frac{1}{2})\sqrt{2k},-\frac{1}{2}\sqrt{2k}>-|-(1+\frac{1}{2})\sqrt{2k},-\frac{1}{2}\sqrt{2k}>)
modH^{Uvk}_{0,0}, \\ \non  
\\ \non 
&& a_{-n_1}(J^+_{-2k}-J^-_{-2k})\bar{a}_{-n_2}
(|\frac{1}{2}\sqrt{2k},\frac{1}{2}\sqrt{2k}> + |-\frac{1}{2}\sqrt{2k},\frac{1}{2}\sqrt{2k}>
-|\frac{1}{2}\sqrt{2k},-\frac{1}{2}\sqrt{2k}>-|-\frac{1}{2}\sqrt{2k},-\frac{1}{2}\sqrt{2k}>) \\ \non  
&=&a_{-n_1} \bar{a}_{-n_2}(|(1+\frac{1}{2})\sqrt{2k},\frac{1}{2}\sqrt{2k}> - |-(1+\frac{1}{2})\sqrt{2k},\frac{1}{2}\sqrt{2k}>
-|(1+\frac{1}{2})\sqrt{2k},-\frac{1}{2}\sqrt{2k}>+|-(1+\frac{1}{2})\sqrt{2k},-\frac{1}{2}\sqrt{2k}>)
modH^{Uvk}_{0,0}.
\label{modind13}
\feqr
\normalsize 
Similarly we can generate the spaces $H^{Uvk}_{0,1}$ and $H^{Uvk}_{1,1}$.  We are now in position to proceed with 
induction.  For the induction we will assume that we have generated the spaces $H^{Uvk}_{s,s'}$ where $0\le s,s'\le p$  
and we will generate the spaces $H^{Uvk}_{p+1,s'}$, $H^{Uvk}_{s,p+1}$ and $H^{Uvk}_{p+1,p+1}$.  To generate the first 
of these spaces we need the relations 
\scriptsize
\beqr 
&& (J^+_{-(2p+2)k}+J^-_{-(2p+2)k}) \Bigl( |(p+\frac{1}{2})\sqrt{2k},(s'+\frac{1}{2})\sqrt{2k}> + 
|-(p+\frac{1}{2})\sqrt{2k},(s'+\frac{1}{2})\sqrt{2k}> \\ \non
&+& |(p+\frac{1}{2})\sqrt{2k},-(s'+\frac{1}{2})\sqrt{2k}>
+|-(p+\frac{1}{2})\sqrt{2k},-(s'+\frac{1}{2})\sqrt{2k}> \Bigr) \\ \non 
&=& \Bigl(|(p+1+\frac{1}{2})\sqrt{2k},(s'+\frac{1}{2})\sqrt{2k}> + |-(p+1+\frac{1}{2})\sqrt{2k},(s'+\frac{1}{2})\sqrt{2k}>
\\ \non 
&+& |(p+1+\frac{1}{2})\sqrt{2k},-(s'+\frac{1}{2})\sqrt{2k}>+|-(p+1+\frac{1}{2})\sqrt{2k},-(s'+\frac{1}{2})\sqrt{2k}> \Bigr)
modH^{Uvk}_{p-1,s'}, \\ \non  
\\ \non 
&& a_{-n}(J^+_{-(2p+2)k}-J^-_{-(2p+2)k}) \Bigl(|(p+\frac{1}{2})\sqrt{2k},(s'+\frac{1}{2})\sqrt{2k}> + 
|-(p+\frac{1}{2})\sqrt{2k},(s'+\frac{1}{2})\sqrt{2k}> \\ \non
&+& |(p+\frac{1}{2})\sqrt{2k},-(s'+\frac{1}{2})\sqrt{2k}>+|-(p+\frac{1}{2})\sqrt{2k},-(s'+\frac{1}{2})\sqrt{2k}> \Bigr) 
\\ \non   
&=&  a_{-n} \Bigl(|(p+1+\frac{1}{2})\sqrt{2k},(s'+\frac{1}{2})\sqrt{2k}> 
- |-(p+1+\frac{1}{2})\sqrt{2k},(s'+\frac{1}{2})\sqrt{2k}> \\ \non
&+& |(p+1+\frac{1}{2})\sqrt{2k},-(s'+\frac{1}{2})\sqrt{2k}>-|-(p+1+\frac{1}{2})\sqrt{2k},-(s'+\frac{1}{2})\sqrt{2k}> \Bigr)
modH^{Uvk}_{p-1,s'}, \\ \non  
\\ \non 
&& (J^+_{-(2p+2)k}+J^-_{-(2p+2)k}) \bar{a}_{-n} \Bigl(|(p+\frac{1}{2})\sqrt{2k},(s'+\frac{1}{2})\sqrt{2k}> + 
|-(p+\frac{1}{2})\sqrt{2k},(s'+\frac{1}{2})\sqrt{2k}> \\ \non
&-& |(p+\frac{1}{2})\sqrt{2k},-(s'+\frac{1}{2})\sqrt{2k}>-|-(p+\frac{1}{2})\sqrt{2k},-(s'+\frac{1}{2})\sqrt{2k}> \Bigr)
 \\ \non 
&=& \bar{a}_{-n} \Bigl(|(p+1+\frac{1}{2})\sqrt{2k},(s'+\frac{1}{2})\sqrt{2k}> + 
|-(p+1+\frac{1}{2})\sqrt{2k},(s'+\frac{1}{2})\sqrt{2k}> \\ \non
&-& |(p+1+\frac{1}{2})\sqrt{2k},-(s'+\frac{1}{2})\sqrt{2k}>-|-(p+1+\frac{1}{2})\sqrt{2k},-(s'+\frac{1}{2})\sqrt{2k}> \Bigr)
modH^{Uvk}_{p-1,s'}, \\ \non  
\\ \non 
&& a_{-n_1}(J^+_{-(2p+2)k}-J^-_{-(2p+2)k})\bar{a}_{-n_2}
\Bigl(|(p+\frac{1}{2})\sqrt{2k},(s'+\frac{1}{2})\sqrt{2k}> + 
|-(p+\frac{1}{2})\sqrt{2k},(s'+\frac{1}{2})\sqrt{2k}> \\ \non
&-& |(p+\frac{1}{2})\sqrt{2k},-(s'+\frac{1}{2})\sqrt{2k}>-|-(p+\frac{1}{2})\sqrt{2k},-(s'+\frac{1}{2})\sqrt{2k}> \Bigr) 
\\ \non 
&=& a_{-n_1} \bar{a}_{-n_2}  \Bigl(|(p+1+\frac{1}{2})\sqrt{2k},(s'+\frac{1}{2})\sqrt{2k}> -
 |-(p+1+\frac{1}{2})\sqrt{2k},(s'+\frac{1}{2})\sqrt{2k}> \\ \non
&-& |(p+1+\frac{1}{2})\sqrt{2k},-(s'+\frac{1}{2})\sqrt{2k}>+|-(p+1+\frac{1}{2})\sqrt{2k},-(s'+\frac{1}{2})\sqrt{2k}> \Bigr)
modH^{Uvk}_{p-1,s'}.   
\label{modind14}
\feqr
\normalsize 
Similarly we generate the other two spaces.  So finally we have generated the whole $H^v_{k/4}$.

\bibliographystyle{plain}

\end{document}